\documentclass[aps,pre,10pt,twocolumn,showkeys]{revtex4-2}

\usepackage[normalem]{ulem}
\usepackage{amsmath,amsfonts}
\usepackage{color,physics,graphicx,bm}
\usepackage{hyperref}

\usepackage[makeroom]{cancel}
\usepackage{mathbbol}
\DeclareSymbolFontAlphabet{\amsmathbb}{AMSb}%

\begin{document}
\title{Tethering effects on first-passage variables of lattice random walks in linear and quadratic focal point potentials}
\author{Debraj Das}
\email{ddas@sissa.it}
\affiliation{SISSA -- International School for Advanced Studies, Via Bonomea 265, 34136 Trieste, Italy}
\affiliation{Istituto dei Sistemi Complessi, Consiglio Nazionale delle Ricerche, via Madonna del Piano 10, 50019 Sesto Fiorentino, Italy}

\author{Luca Giuggioli}
\email{Luca.Giuggioli@bristol.ac.uk}
\affiliation{School of Engineering Mathematics and Technology,
University of Bristol, Bristol BS8 1TW, United Kingdom}

\date{\today}

\begin{abstract}
Diffusion in a confining potential offers a minimal setting to understand the interplay between random motion and deterministic forces driving a particle towards a focal point or potential minimum. In continuous space and time, two extensively studied examples are Brownian motion in a linear (V--shaped) or a quadratic (U--shaped) potential. The deterministic bias towards the minimum is represented, respectively, by a constant force for the former and by an elastic restoring force that increases proportionally with distance for the latter. Surprisingly, unlike Brownian walks, random walks under focal point potentials in discrete space and time have received little attention. Here, we bridge this gap by analysing the dynamics of lattice random walkers in the presence of a V--shaped potential, both in a finite and an infinite spatial domain, and a finite U--shaped potential. For the V--potential in unbounded space, we find the generating function of the occupation probability and analyse the time dependence of the mean number of distinct sites visited, demonstrating that its long-time growth is logarithmic. We also study the first-passage probability and show that its mean may display a minimum as a function of bias strength, depending on the location of the initial and target sites relative to the focal point. Qualitatively similar dependencies in the first-passage probability and its mean appear for the finite U--potential. As a comparative analysis to the U--potential, we construct the bounded V--potential and superimpose in both cases a resetting process, in which the walker returns at random times to a site distinct from the focal point with some probability. We quantify the different effects of resetting on the steady-state probability and the first-passage dynamics in the two cases, and show a motion-limited regime emerges even for relatively moderate resetting probabilities.
\end{abstract}


\maketitle

\section{Introduction}

Random walks are among the simplest stochastic processes, yet they represent a very useful laboratory to capture essential aspects of transport, diffusion, and stochastic search in disordered or structured media.
Since the early formulations by Pearson~\cite{pearson1905problem} and Polya~\cite{polya_quelques_1918,polya_uber_1921}, lattice random walks (LRWs), that is, random walks in discrete time and discrete space, have served as essential models of transport across physics, chemistry, and biology. 
Classic monographs by Weiss~\cite{weiss1994aspects} and Hughes~\cite{huges1995random} established LRWs as fundamental to study diffusion and interactions on discrete geometries. 
Among the various observables associated with stochastic motion, the first-passage time—the time at which a walker first reaches a designated target—plays a central role.
As many events of interest, such as arrival, escape, and binding, are governed not by the full trajectory but by the time at which a threshold is first crossed, the theory of  first-passage (FP) processes  has become central to a vast amount of problems in chemical kinetics, biophysics, ecology, and finance~\cite{redner_guide_2001}. Seminal works by Montroll showed how boundary conditions and lattice structure strongly shape FP distributions~\cite{ montroll_random_1965}, and recent developments~\cite{giuggioli_exact_2020} have reinvigorated interest in LRWs due to their applicability to transport problems on networks~\cite{noh2004random,marrisetal2025}, search in heterogeneous media~\cite{sarvaharman_particle-environment_2023, das_dynamics_2023, giuggioli2024multi}, dynamics of random encounters~\cite{riascossanders2021,giuggioli_spatio-temporal_2022,das_misconceptions_2023}, persistent walks~\cite{marris_persistent_2024, sarmiento_first-passage-time_2025}, and quantum measurement engines~\cite{lea2025rnt}.

A current frontier concerns random walks evolving in non-trivial environments, e.g., heterogeneous lattices, space with complex topologies such as networks and random landscapes. 
However, many physical and biological systems are naturally influenced by confining potentials, which create restoring forces that bias the walker deterministically towards a focal point or a particular region of space. 
In continuous space, linear \cite{touchetteetal2010,chaseetal2016,giuggioli_comparison_2019}  and harmonic potentials~\cite{Uhlenbeck_1930, Clercx1992, mancois_two_2018}  are textbook examples in the theory of diffusion under force fields~\cite{risken_fokker-planck_1996, zwanzig_nonequilibrium_2001,goel_stochastic_1974}.
On a lattice, the discrete analogues are, respectively, V--shaped and U--shaped (elastic) potentials. In the former, the potential ``energy'' increases linearly with distance from a central point, alternatively with a constant bias towards the focal point, while in the latter, the restoring force increases proportionally to the distance as one moves away from the energy minimum. 
These discrete potentials are not merely mathematical constructs; they provide minimal models of confinement on a lattice while preserving full analytical tractability. As real physical and biological systems are replete with heterogeneities that attract or bias motion towards special areas, LRWs in focal point potentials offer a controlled framework to investigate how deterministic confinement influences transport characteristics, spreading processes, and first-passage statistics in discrete geometries.

Given the simplicity and yet fundamental nature of the process of diffusion within a concave potential, i.e., possessing a minimum, it would have been natural to expect a large number of analytical studies on the topic. 
Surprisingly, and contrary to the literature with continuous variables, research on LRWs within a confining potential has been rare.
Indeed, recent work on confined continuous processes has revealed nontrivial scaling structures in first-passage statistics, including distinct finite-size and infinite-system regimes~\cite{baravi_first_2025,baravi_solutions_2025}, underscoring the richness of confined dynamics.
While analytic studies of the spatio-temporal dynamics of externally biased LRWs have been explored for finite, semi-infinite, and infinite domains \cite{montroll1967,khantha_reflection_1985,godoy_reflection_1992,sarvaharman_closed-form_2020}, similar level of interest in analytical developments for the V-- and U--shaped potentials has not been forthcoming, with the only exception being the work by Kac in 1947 \cite{kac_random_1947} on the dynamics of the occupation probability in the bounded elastic potential. 
However, even for the elastic potential, a comprehensive
spatio-temporal description, particularly including first-passage properties, has remained unavailable.
In this work, we bridge this gap by developing an exact analytical framework for the dynamics and first-passage statistics of LRWs in one-dimensional V-- and U--shaped  potentials, the former both for unbounded and bounded domains, while the latter only for a bounded domain.

The analytical form of the LRW occupation probability allows us to link our findings to another topic of growing interest, stochastic resetting, whereby  a walker is intermittently returned to a prescribed location~\cite{evans_stochastic_2020,gupta_stochastic_2022}.
In continuous time, resetting has been shown to modify first-passage times, splitting probabilities, and search optimality in bounded chains~\cite{christophorov_resetting_2022,christophorov_continuous_2024} and $d$-dimensional hypercubic lattice~\cite{hartmann_diffusion_2025}. In discrete time,
recent findings on the dynamics of LRWs in a spatially structured domain point to the appearance of a motion limited regime for strong enough resetting probability \cite{barbinigiuggioli2024}, raising new questions about the relative importance of confinement and resetting in shaping first-passage statistics.
As resetting is known to optimise target search and alter FP scaling when space is homogeneous, we investigate how search optimality is affected by the strength of the spatial confinement and the resetting probability. Furthermore, we analyse the temporal growth of a quantity directly related to the first-passage probability, namely the number of distinct sites visited. 

The paper is structured as follows. In Section~\ref{sec:vpot}, we introduce the Master equation describing the occupation probability of an LRW moving in an unbounded V--shaped potential, and solve it exactly to derive the so-called propagator or Green's function. We use it to study the statistics of first-passage  to a single target and of the average number of distinct sites visited, and then construct semi-bounded and fully bounded propagators. Section~\ref{sec:upot} treats the LRW under a finite elastic U--shaped potential, deriving propagators and FP distributions and highlighting differences from the V--shaped case. Section~\ref{sec:reset} studies confined walks with resetting, including FP distributions and the mean number of visited sites. Finally, Section~\ref{sec:discus} summarises our findings and discusses the broader implications of our results, while Appendices provide detailed derivations.

\section{The V--shaped potential}
\label{sec:vpot}

\subsection{The model}

\begin{figure}[t]
\centering
\includegraphics[scale=1]{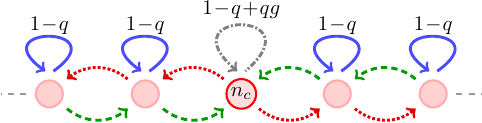}
\caption{Schematic diagram of transition probabilities for a walker under the V--shaped potential centred at the focal site $n_c$ in unbounded space. The red (dotted) and green (dashed) arrows denote the constant probabilities $q(1-g)/2$ and $q(1+g)/2$, respectively, and thus make the potential symmetric about $n_c$.}
\label{fig:deb_vpot}
\end{figure}

Let us consider a 1D dynamics of a walker under a V--shaped potential centred at $n_c$ such that it experiences a constant bias that pulls it towards $n_c$. For such a confinement, the effective ``energy'' increases linearly with the distance from the centre, given by a potential $\mathcal V (n) = \kappa\,|n - n_c|$, with $\kappa>0$ setting the overall strength. The corresponding discrete force,
$\mathfrak F(n) \propto -\,\mathrm{sgn}(n - n_c)$, points towards $n_c$ for all $n\neq n_c$, reversing discontinuously across the centre.
To incorporate this force into an LRW framework, we assign asymmetric hopping probabilities that depend on the sign of $(n-n_{c})$. Introducing a parameter $g \in [0,1)$ that controls the steepness of the potential, a diffusivity parameter \(q\in[0,1]\), 
and  denoting the probability to find the walker on site $n$ at time $t$ by $Q(n,t)$, one may write the following Master equation:
\begin{align}
Q(n,t+1) &=  [ 1-q(1 - g \delta_{n,n_c} ) ] Q(n,t) \nonumber \\
&\hskip-20pt +\frac{q}{2} [ 1 + g \qty( \delta_{n,n_c} - {\mathrm{sgn}} (n-n_c) ) ] Q(n-1,t)  \nonumber \\
&\hskip-20pt +\frac{q}{2} [ 1 + g \qty( \delta_{n,n_c} + {\mathrm{sgn}} (n-n_c) ) ] Q(n+1,t)  .
\label{eq:deb_vpot_master_compact}
\end{align}
The walker, when at $n \neq n_c$, hops towards the centre with probability $q(1+g)/2$, hops away from it with probability $q(1-g)/2$, and remains at the same site with probability $(1-q)$. 
At $n=n_c$, the dynamics is symmetric, with equal hopping probability $q(1-g)/2$ and a rest probability $(1-q+qg)$,  see Fig.~\ref{fig:deb_vpot}.
As $g$ grows, the potential becomes steeper, and the walker is increasingly drawn towards the centre.
For $g=0$, the dynamics reduces to that of a lazy symmetric random walk with hopping probability $q/2$ and rest probability $(1-q)$ \cite{giuggioli_exact_2020}.

\subsection{Unbounded propagator}

The presence of the focal site $n_c$, at which the direction of drift reverses, renders Eq.~\eqref{eq:deb_vpot_master_compact} substantially more intricate than the Master equation of a conventional biased random walk. 
In standard biased diffusion, the drift maintains a uniform direction across the lattice, allowing closed--form solutions through classical generating--function and Fourier techniques~\cite{sarvaharman_closed-form_2020}. 
In contrast, the drift in Eq.~\eqref{eq:deb_vpot_master_compact} changes sign exactly at $n_c$, producing a discontinuity in the effective force landscape that couples the dynamics on the two sides of the lattice in a non-trivial manner. 
A practical way to interpret dynamics~\eqref{eq:deb_vpot_master_compact} is within the broader framework of heterogeneous random walks, in which the medium consists of two regions with distinct transport properties separated by an interface~\cite{das_dynamics_2023}.
In such heterogeneous settings, the form of the master equation depends sensitively on whether the interface lies \emph{between} lattice sites (Type A) or exactly \emph{on} a lattice site (Type B)~\cite{das_dynamics_2023}. 
The present problem corresponds to a special case of a Type~B interface, where the two media share the same diffusivity but possess drifts of equal magnitude and opposite direction, both pointing towards the interface at $n_c$.

By mapping the parameters in Eq.~\eqref{eq:deb_vpot_master_compact} to the Type B interface problem [see Appendix~\ref{sec:Vpot-prop-app}], one obtains an exact solution of the Master equation, that is the so-called Green's function or propagator $Q(n,t | n_0)$ for a walker initially at site $n_0$, in terms of its generating function $\widetilde{Q}(n,z|n_0) \equiv \sum_{t=0}^{\infty} z^t Q(n,t|n_0)$, as
\begin{align}
\widetilde{Q} (n,z|n_0) &= 
 \frac{f^{-|n_c-n_0|}  \xi^{|n-n_0|} }{ \chi  (\chi + z q g) }  \big( \chi f^{\alpha / 2}  +  z q g  \xi^{\alpha} \big)   ,\label{eq:GF_inZ_VPot_gen_sol}
\end{align}
where we have
\begin{align}
& \alpha \equiv |n_c-n|+|n_c-n_0|-|n-n_0| \, , \\
& \chi(z)  \equiv  -z q  g + \qty[1-z(1-q)] \sqrt{1-\beta_{+}\beta_{-}}  \, , \label{eq:chi-def}\\
& f \equiv \frac{1\!-\!g}{1\!+\!g}, \, \beta_{\pm} (z) \equiv \frac{z q(1 \! \pm \!  g)}{1\!-\!z(1\!-\!q)}  , \, \xi(z) \equiv \frac{1\!-\!\sqrt{1\!-\!\beta_{+} \beta_{-}}}{\beta_{+}} .  \label{eq:xi-def} 
\end{align}
Note that when $n_c$ is between $n$ and $n_0$, one has $\alpha = 0$ and $\widetilde{Q} (n,z|n_0) = f^{-|n_c-n_0|} \, \xi^{|n-n_0|} / \chi$.
In the absence of bias, i.e. $g=0~(f=1)$, the focal point and the V--potential cease to exist, and Eq.~\ref{eq:GF_inZ_VPot_gen_sol} reduces to the lazy lattice-walk propagator \cite{giuggioli_exact_2020} $\widetilde{Q}(n,z|n_0) = \xi^{|n-n_0|} / [\sqrt{1-\beta^2 } \, \big( 1-z+z q \big)]   $
with $ \beta = z q / (1-z+zq) $ and $\xi = 1/\beta - \sqrt{1/\beta^2-1}$.

In the limit $t \to \infty$, using the final value theorem for $z$-transform, the steady-state probability $Q^{\mathrm{ss}}(n) \equiv  \lim_{z \to 1} [(1 - z) \widetilde{Q}(n,z|n_0)]$ is obtained as [see Appendix~\ref{sec:Vpot-prop-app}]
\begin{align}
Q^{\mathrm{ss}}(n)  = g \, f^{|n-n_c|} ,
\label{eq:VPot_ss}
\end{align}
The steady state parameter dependence leads to the expected mean, $\expval{n} = \sum_{n=-\infty}^{\infty} n Q^{\mathrm{ss}}(n) = n_c$, and variance ${\mathrm{Var}}(n) = \expval{n^2} - \expval{n}^2 = (1/g^2-1)/2 $, which decreases as the potential gets steeper, i.e. with increasing $g$, while it becomes unbounded in the absence of a potential, i.e. $g \to 0$.
From Eq.~(\ref{eq:VPot_ss}) it is also noteworthy to point out that $Q^{\mathrm{ss}}(n)$   does not depend on the diffusivity $q$, something  markedly different from the continuous counterpart, which is explicitly dependent on the ratio of the potential strength to the diffusion constant~\cite{chaseetal2016}.
Similarly to the description with continuous variables, where the diffusion constant rescales the dynamics, the diffusivity parameter affects the time-dependence of the discrete occupation probability, but differently from the continuous case, here the potential strength and diffusivity are not independent because of probability conservation constraints. The potential strength is in fact linearly dependent on the diffusivity. As the ratio of the potential strength divided by the diffusivity eliminates the $q$-dependence, this is why the occupation probability at steady state only depends on the bias parameter $g$ (and the lattice site position relative to the potential minimum).

\begin{figure}[t]
    \centering
    \includegraphics[scale=1]{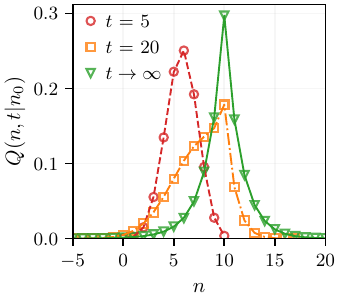}
    \caption{Propagator $Q(n, t|n_0)$ for  dynamics~\eqref{eq:deb_vpot_master_compact} under a V--potential centred at site $n_c=10$ in unbounded spatial domain. The walker starts from site $n_0=5$ with $q=0.5$ and $g=0.3$. The points are obtained from $5\times10^5$ stochastic realizations of dynamics~\eqref{eq:deb_vpot_master_compact}. The broken lines are obtained by numerically inverting Eq.~\eqref{eq:GF_inZ_VPot_gen_sol}, while the continuous line in green denotes the steady-state given in Eq.~\eqref{eq:VPot_ss}.}
    \label{fig:vpot_prop}
\end{figure}

As the analytical time dependence of Eq.~\eqref{eq:GF_inZ_VPot_gen_sol} is the sum of four time dependent terms with each term represented by concatenated function of  convolutions (see Appendix \ref{sec:Vpot-prop-app}), the plot of $Q(n,t | n_{0})$ have been more conveniently obtained by numerically inverting~\cite{abate_numerical_1992} Eq.~\eqref{eq:GF_inZ_VPot_gen_sol} and displayed in Fig.~\ref{fig:vpot_prop}. 
One may see that at early and intermediate times ($t=5,20$) the probability profile shifts towards the centre $n_{c}$, because of the inward pull induced by the drift,  reaching the steady state, which displays a cusp at $n_c$ as per $|n-n_c|$ dependence in Eq. (\ref{eq:VPot_ss}).

\subsection{First-passage statistics}

The generating function of the first-passage time probability to reach a target site $n$ starting from $n_0$ is given by the renewal relation $\widetilde{F}(n,z|n_0) = \widetilde{Q}(n,z|n_0)/\widetilde{Q}(n,z|n)$, yielding
\begin{align}
    \widetilde{F}(n,z|n_0) =  \frac{f^{|n_c-n|-|n_c-n_0|}  \xi^{|n-n_0|}   ( \chi f^{\alpha / 2}  +  z q g  \xi^{\alpha} ) }{  \chi f^{|n_c-n|}  +  z q g \, \xi^{2|n_c-n|}  }   .
    \label{eq:FPT_inZ_VPot}
\end{align}
 Using the relation $\expval{T} = \lim_{z\to1} \partial{\widetilde{F}(n,z|n_0)}/\partial{z}$, with $\expval{T}$ representing the mean first-passage time to reach $n$ from $n_0$, we obtain
\begin{align}
    \expval{T} = \frac{g | n_0-n_c| -g | n-n_c| +f^{-| n-n_c| }  -f^{-{\alpha}/{2}} }{g^2 q}  ,
    \label{eq:MFPT_Vpot}
\end{align}
which diverges as expected for a lazy walker in the limit $g\to 0$. The form of Eq.~\eqref{eq:MFPT_Vpot} lends itself to some interesting observations.
When the target site $n$ lies between the initial position $n_0$ and the potential minimum at $n_c$, i.e., $n_0 < n \leq n_c$ (or symmetrically $n_c \leq n < n_0$), one has $\expval{T} = \frac{|n-n_0|}{q g}$, i.e., the mean first-passage time simply equals the ballistic travel time from $n_0$ to $n$ at effective speed $q g$, something noted also for the corresponding dynamics with continuous variables~\cite{giuggioli_comparison_2019}.
For $n>n_0>n_c$ (or symmetrically $n<n_0<n_c$), we obtain $\expval{T}_{n_0 \to n} = \expval{T}_{n_c \to n} - \expval{T}_{n_c \to n_0}$, where the suffix $i\to j$ in $\expval{T}_{i\to j}$ denotes from site $i$ to $j$. In other words, in this case, the mean first-passage time is the average time the walker would need to get to $n$ from $n_c$ minus the contribution to get to $n_0$ from $n_c$.
On the other hand, if $n$ and $n_0$ belong to the opposite halves of the potential, i.e., for $n_0<n_c<n$ (or symmetrically $n_0>n_c>n$), one has $\expval{T}_{n_0 \to n} = \expval{T}_{n_0 \to n_c} + \expval{T}_{n_c \to n}$, i.e., the contribution of the sum of two terms, namely the average time to get to the bottom of the potential and from there (the average time) to reach $n$.

The second moment of the first-passage time can be obtained from the generating function via $\expval{T^2} = \expval{T} + \lim_{z\to1} \partial^2{\widetilde{F}(n,z|n_0)}/\partial{z}^2$, which allows to compute the corresponding variance as
\begin{align}
    \mathrm{Var}(T) &= \frac{1}{g^4 q^2} \Big[  g ( 1-g^2 q ) \big( | n_c-n_0| -| n-n_c| \big) \nonumber \\
    &\hskip40pt +\Phi(| n-n_c| ) -\Phi(\alpha/2) \Big] ,
    \label{eq:VarFPT_Vpot}
\end{align}
where the function $\Phi(x)$ is defined by $\Phi(x) \equiv f^{-2 x}-f^{-x} \left[g^2 (q+1)+4 g x-1\right]$.

\begin{figure}[t]
    \centering
    \includegraphics[scale=0.72]{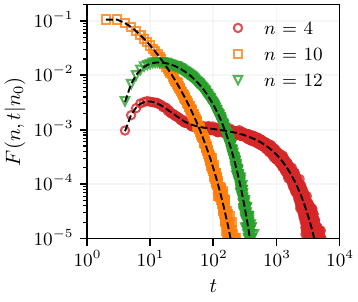}\hskip-2pt
    \includegraphics[scale=0.72]{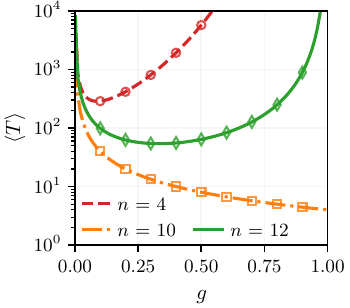}\\[0.2ex]
    \includegraphics[scale=0.73]{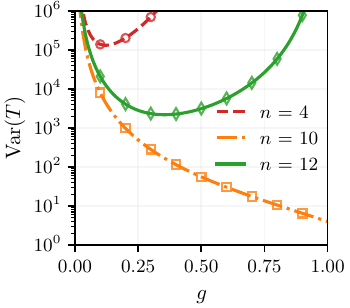}\hskip1pt
    \includegraphics[scale=0.73]{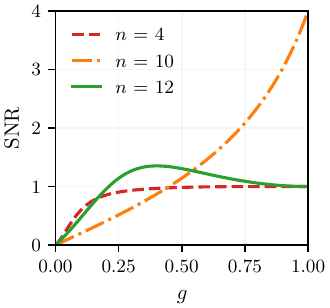}
    \caption{First-passage walker statistics in a V--potential centred at site $n_c=10$ in unbounded space. The walker starts from site $n_0=8$ and reaches a target at $n$ with $q=0.5$. In the plots, the points are obtained from $5\times10^5$ stochastic realizations of dynamics~\eqref{eq:deb_vpot_master_compact}, while the lines denote analytical results. Top left: First-passage probability $F(n, t|n_0)$ as a function of time $t$ for different targets $n$ with $g=0.3$. The dashed lines in black are obtained by numerically inverting~\cite{abate_numerical_1992} Eq.~\eqref{eq:FPT_inZ_VPot}. Top right: Mean first-passage time $\expval{T}$ from Eq.~\eqref{eq:MFPT_Vpot}. Bottom left: Variance of the first-passage time from Eq.~\eqref{eq:VarFPT_Vpot}. Bottom right: The signal-to-noise ratio obtained using the relation $\mathrm{SNR}=\expval{T}^2/\mathrm{Var}(T)$. Note that for $g=0$ the dynamics reduce to that of a symmetric random walk, leading to diverging mean and variance for all target sites.}
    \label{fig:vpot_fpt}
\end{figure}

Figure~\ref{fig:vpot_fpt} summarises the first-passage properties of the V--shaped potential in an unbounded domain.  
The top--left panel of Fig.~\ref{fig:vpot_fpt} shows the first-passage probability $F(n,t|n_0)$ as a function of time $t$ for three different target locations.  
The qualitative shapes of the three curves directly reflect the relative positions of the starting and target location, and the focal site.
For the target at $n=10$, which coincides with the centre of the potential, the drift is always directed towards the target.  
Consequently, the first-passage distribution is sharply peaked at short times and decays rapidly, indicating that the arrival time is both fast and highly predictable.
For the target at $n=12$, which lies to the right of both $n_0$ and $n_c$, the walker is initially pushed towards the target by the drift (from $8$ to $10$), but once it crosses the centre the drift tends to pull it back.  
The resulting distribution therefore peaks at intermediate times and its broader tail reflects the competition between drift towards and drift away from the target.
For the target at $n=4$, the walker must move entirely against the bias in order to reach the target.  
Early arrivals are thus strongly suppressed, and many trajectories spend a long time wandering near the focal site before making a rare excursion to the left.  
This produces the broad, slowly decaying first-passage distribution seen for $n=4$, whose tail is considerably
heavier than in the $n=10$ and $n=12$ cases.

The top--right panel of Fig.~\ref{fig:vpot_fpt} shows the mean first-passage time $\langle T \rangle$ for different target locations as a
function of the bias strength $g$.  
For $n=10$, the drift always points directly towards the target: increasing $g$ strengthens this inward pull and leads to a monotonic decrease of $\langle T \rangle$.
For $n=12$, the drift helps the walker reach the centre from $n_0=8$, but opposes the final motion from $n_c=10$ to the target. 
This competition produces a non-monotonic behaviour: $\langle T \rangle$ decreases at small $g$ when the  assistance to the centre dominates, but increases at large $g$ once the drift strongly traps the walker near the centre.
For $n=4$, the drift always acts against the required motion, making leftward excursions increasingly unlikely as $g$ grows: $\langle T \rangle$ increases sharply with $g$, as one may evince from the sharp increase on the vertical logarithmic scale. These qualitative features of the mean are also displayed in the variance, which is plotted in the bottom--left panel.

The bottom--right panel of Fig.~\ref{fig:vpot_fpt} shows the signal-to-noise ratio, $\mathrm{SNR}=\langle T\rangle^{2}/\mathrm{Var}(T)$, which quantifies how predictable the first-passage time is: a larger SNR indicates smaller fluctuations relative to the mean first-passage time. When the target is the centre of the potential ($n=n_c=10$), the drift always assists motion towards it, so increasing the bias $g$ reduces both the mean and the fluctuations.  
Because the variance decreases  faster than the square of the mean, the SNR grows monotonically and becomes largest when the drift is strongest.
For targets that are not fully aligned with the downhill direction of the 
potential (e.g., $n=12,4$), the role of the bias is more subtle. In the strong-bias limit $g\to 1$, the potential becomes so steep that the walker approaches the centre  $n_{c}$ very quickly and any displacement away from it is pulled back almost immediately.
As a result, the trajectory consists of many short excursions away from the centre, each of which is nearly identical and has an approximately constant probability of reaching the target before being pulled back to $n_c$.  
These excursions act as independent attempts with fixed success probability, and thus the long-time statistics of the first-passage process become effectively \emph{memoryless} (the process has no memory of how many failures have occurred before). The first-passage distribution therefore develops an exponential tail, for which 
$\mathrm{Var}(T)\approx \langle T\rangle^{2}$, causing the SNR to approach~1  as $g\to 1$.
For a target located to the right of the centre ($n=12$), a weak bias initially 
helps the walker reach the target, but a stronger bias increasingly restricts 
exploration by trapping the walker near $n_c=10$.
This competition leads to a non-monotonic behaviour of the SNR, which shows a maximum value greater than 1 around an intermediate value of $g$.
In contrast, for a target to the left 
($n=4$), the bias always opposes the motion towards the target, so the SNR increases monotonically and saturates rapidly to 1 as $g$ increases.

\subsection{Average number of distinct sites visited}

A useful observable that captures the extent with which the probability spreads in space is the mean number of distinct lattice sites visited, $\langle N(t)\rangle$, during the course of a $t$--step walk. 
The probability that a given site $n$ is visited at least once during the time window $t$ is given by the cumulative probability $\sum_{t'=1}^{t} F (n,t'|n_0)$ with $F (n,t'|n_0)$ being the first-passage probability~\cite{prigogine_random_1982}.
The average number of distinct sites visited within time $t$ is given by~\cite{prigogine_random_1982}
$ \expval{N(t)} = \sum_{n} \sum_{t'=1}^{t} F (n,t'|n_0)$.
For a nearest-neighbour random walk on an unbounded 1D lattice without external forces, the walker typically explores a region of width $\sqrt{t}$, leading to the well-known scaling $\expval{N(t)} \sim \sqrt{t}$~\cite{prigogine_random_1982,dayan_number_1992}.  
In the presence of a confining potential, even if the space is unbounded, $\expval{N(t)}$ can still grow without bound at long times. 
This behaviour contrasts with that of the mean square displacement, which instead saturates even when the tethering force towards a focal point increases very slowly 
with distance, for instance logarithmically~\cite{giuggiolietal2006}.
The V--potential represents one example whereby the confinement induced by the potential is not strong enough to make $\langle N(t)\rangle$ saturate as $t\to \infty$.
To determine how $\langle N(t)\rangle$ grows for large $t$, we invert numerically its generating function~\cite{ biroli_number_2022}
\begin{align}
\langle \widetilde{N}(z) \rangle=\frac{1}{1-z}\sum_{n}\widetilde{F}(n,z|n_0),
\label{eq:MNDSV}
\end{align}
with $\widetilde{F}(n,z|n_0)$ defined in Eq.~\eqref{eq:FPT_inZ_VPot}.

\begin{figure}[t]
    \centering
    \includegraphics[scale=0.74]{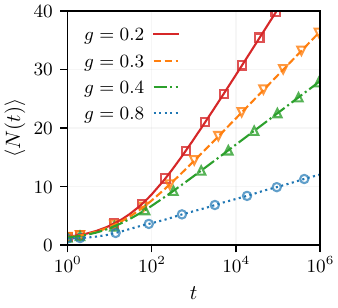}
    \includegraphics[scale=0.74]{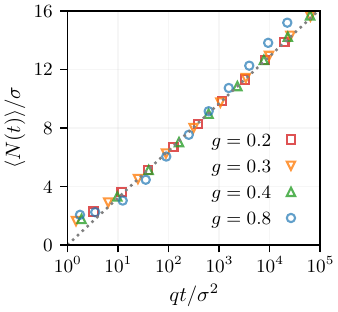}
    \caption{Average number of distinct sites visited $\expval{N(t)}$ by a walker moving within the V--potential with different values of $g$ in unbounded domain. The walker starts from the centre of the potential $(n_0=n_c=0)$ with diffusivity $q=0.5$. The left panel shows $\expval{N(t)}$ vs $t$, while the right panel shows the scaling law. Points are obtained from $2000$ stochastic realizations. The lines in the left panel are obtained by numerically inverting Eq.~\eqref{eq:MNDSV}, where the infinite spatial sum is truncated to the range $-100 \le n \le 100$, while the dotted straight line in the right panel is shown as a guide to the eye. }
    \label{fig:Nt-vs-t-vpot-unbou}
\end{figure}
The left panel of Fig.~\ref{fig:Nt-vs-t-vpot-unbou} shows the mean number of distinct sites visited for the V--potential in unbounded space. 
At long times the bias pulls the walker towards the focal point $n_c$, and most revisits occur within a typical confinement length $\sigma \equiv \sqrt{{\mathrm{Var}}(n)} = \sqrt{(1/g^2-1)/2}$.
Although this spatial concentration strongly suppresses the discovery of new locations, it does not halt exploration: $\langle N(t)\rangle$ continues to grow unboundedly, albeit much more slowly than a free lattice walker. 
Our results are consistent with a logarithmic increase of $\langle N(t)\rangle$ in time at long times.

The right panel of Fig.~\ref{fig:Nt-vs-t-vpot-unbou} shows that, upon rescaling time as $qt/\sigma^{2}$ and the vertical axis as  $\langle N(t)\rangle/\sigma$, the curves for different values of $g$ approximately collapse onto a common scaling function.
This demonstrates that, for moderate bias strengths, the dependence on $g$ enters primarily  through the confinement length $\sigma$, which governs the effective spatial exploration window. 
For strong confinement (e.g., $g=0.8$ for which $\sigma \approx 0.53 < 1$), systematic deviations from the scaling function are observed, reflecting the distinct microscopic exploration dynamics of a walker that remains highly localized near the focal site.

\subsection{Reflecting domain}

Analytic tractability is preserved also when the walker moves in a semi-bounded or fully bounded domain. Here we derive analytic expressions when the walker in the V--shaped potential with  the minimum of the potential at site $n_c = R$ is restricted to either a semi-infinite domain, $n\geq 0$, with a reflective boundary at $n=0$, or a finite interval, $n\in[0,2R]$ with reflecting boundaries at sites $0$ and $2R$. 
In other words, we consider hard walls at one or both ends of the domain, where  the walker may either remain still  or move towards $n_c$.

In random walk problems in homogeneous space, reflecting boundaries can often be treated using the classical method of images.  
In heterogeneous systems, however, this approach breaks down because the transition rules vary across the lattice and the required symmetry  to construct image solutions is absent.   
To address these scenarios, we use Montroll's defect technique~\cite{montroll_effect_1955}, recently extended to heterogeneous spaces, in which a boundary is represented by a local defect that alters the transition probabilities between pairs of sites~\cite{sarvaharman_particle-environment_2023,das_dynamics_2023}. 
The method allows to conserve probability, altering the dynamics only at the heterogeneity and without affecting any of the dynamics away from it. 
This provides a systematic and symmetry–independent method for incorporating reflecting or partially reflecting boundary conditions, and more generally any inert (probability preserving) spatial heterogeneities \cite{sarvaharman_particle-environment_2023}.

\subsubsection{Semi-bounded propagator}

To construct the full propagator in the bounded domain, it is convenient to impose the reflecting conditions one boundary at a time, following the general approach in \cite{das_dynamics_2023}.
We begin by introducing a single reflecting defect at site $0$; this yields the \emph{semi-bounded} (left-bounded) propagator, whereby when the walker is at site $0$, its next move consists of either staying at $0$ with probability $1 - q(1+g)/2$ or moving to site $1$ with probability $q(1+g)/2$.
Considering this modification, the left semi-bounded propagator, $\widetilde L(n,z | n_0)$, can be exactly expressed in terms of the unbounded propagator $\widetilde Q(n,z | n_0)$ given in Eq.~\eqref{eq:GF_inZ_VPot_gen_sol} as~\cite{das_dynamics_2023}
\begin{align}
\widetilde L(n,z | n_0)
&= \widetilde Q(n,z | n_0) + \widetilde Q(0,z | n_0) \nonumber \\
&\hskip-30pt \times \frac{ z q (1-g)  \big[ \widetilde Q(n,z | 0) - \widetilde Q(n,z | -1) \big] }{ 2 - z q (1-g) \big[ \widetilde Q(0,z | 0) - \widetilde Q(0,z |-1) \big] },
\label{eq:Vpot-left-bound-prop}
\end{align}
where $n,n_0 \geq 0$,
with steady state, $L^{\rm ss}(n) \equiv  \lim_{z\to1}(1-z)\widetilde L(n,z| n_0)$, given by [see appendix~\ref{sec:Vpot-refl-prop-app}]
\begin{align}
L^{\rm ss}(n) = \frac{2 g f^{| n-n_c| }}{2-(1-g) f^{ n_c}} .
\label{eq:Vpot-semi-ss}
\end{align}
The spatially asymmetric shape of $L^{\rm ss}(n)$ is manifested in the explicit bias and $n_c$ dependence of the mean and variance of the walker's position given, respectively, by
\begin{align}
    \expval{n} &= \frac{1+f}{1+f-f^{n_c+1}} \qty(n_c + \frac{f^{n_c+1}}{1-f^2} ), \quad \text{and}\\
    {\mathrm{Var}}(n) &= \frac{f}{4g^2(1+f-f^{n_c+1})^2} \big[ 8 + f^{n_c} \{ (1-g)^2 f^{n_c}\nonumber \\
    &\hskip25pt  -4 g n_c ((1-f) n_c+2) -8 +4g \} \big] .
\end{align}

\subsubsection{Fully-bounded propagator}
\label{sec:vpot-fully-bounded}

To introduce a reflecting site to the right at $2R$, we use the semi-bounded propagator $\widetilde L(n,z | n_0)$  and treat the reflecting site at $2R$ as a defect. 
When the walker is at $2R$, it remains there with probability $1 - q(1+g)/2$ or moves to  $2R-1$ with probability $q(1+g)/2$. 
Using the same approach as in (\ref{eq:Vpot-left-bound-prop}), the fully-bounded propagator $\widetilde Q^{\mathrm{ref}}(n,z | n_0)$ is given by
\begin{align}
    \widetilde Q^{\mathrm{ref}}(n,z | n_0)
    &= \widetilde L(n,z | n_0) +   \widetilde L(2R,z | n_0) \nonumber \\
    &\hskip-50pt \times \frac{
    z q (1-g)
    \big[ \widetilde L(n,z | 2R) - \widetilde L(n,z | 2R+1) \big] }
    {2 - z q (1-g)
      \big[ \widetilde L(2R,z | 2R) - \widetilde L(2R,z | 2R+1) \big] },
\label{eq:Vpot-fully-bound-prop}
\end{align}
with $n,n_0 \in [0,2R].$
The steady-state in this case is given by [see appendix~\ref{sec:Vpot-refl-prop-app}]
\begin{align}
Q^{\mathrm{ref}}_{\mathrm{ss}}(n)  = \frac{g f^{| n-R| }}{1 -  (1-g) f^R} , 
\label{eq:Vpot-refl-full-ss}
\end{align}
which is symmetric around $n=R$, with the mean position of the walker  being $\expval{n} = \sum_{n=0}^{2R} n Q^{\mathrm{ref}}_{\mathrm{ss}}(n) = R$, while the corresponding variance is given by
\begin{align}
    {\mathrm{Var}}(n) = \frac{ f [ 1+g-f^R (1+g+2 g R (1+g+g R) ) ] }{ g^2 [1+f-2 f^{R+1}]}  .
    \label{eq:vpot-fully-var-n}
\end{align}
Remarkably, apart from a domain-size-dependent normalization, the fully bounded steady state $Q^{\mathrm{ref}}_{\mathrm{ss}}(n)$ is identical in form to the unbounded steady state of Eq.~\eqref{eq:VPot_ss}, retaining the characteristic cusp at the centre of the V--shaped potential.

\subsubsection{First-passage statistics in fully-bounded domain}

The propagator $\widetilde Q^{\mathrm{ref}}(n,z | n_0)$ gives us access to the generating function of the first-passage probability $\widetilde F^{\mathrm{ref}}(n, z | n_0) = {\widetilde Q^{\mathrm{ref}}(n, z | n_0) } / { \widetilde Q^{\mathrm{ref}}(n, z | n) } $, from which one may derive the mean first-passage time for the walker to reach $n$ starting from $n_0$ [see appendix~\ref{sec:Vpot-refl-prop-app}]
\begin{align}
    \expval{T} = \frac{| n_0-R| -| n-R| }{g q} + \frac{f^{-| n-R|} \big[ \theta (n)-\theta (n_0) \big]}{(1-f) g q} ,
    \label{eq:vpot-fully-mfpt}
\end{align}
where we have $\theta(x) \equiv f^{\frac{1}{2} (| n+n_0-R-x| -| n_0-R| )} [ (1+f^{R-x}) f^{\frac{1}{2} (| n-R| +R+x+2)}+ ( 1+f -2 f^{R+1} ) f^{\frac{| n-x| }{2}}  ] $. 
When $R\leq n<n_0$ or $n_0<n\leq R$ from Eq. (\ref{eq:vpot-fully-mfpt}) one can see that the first term represents the time it takes the mean position of the walker, initially at $n_0$, to reach $n$ moving with speed  $q g$.

\section{The elastic U--shaped potential}
\label{sec:upot}

\subsection{The model}

We now consider the discrete lattice analogue of a particle evolving under a quadratic potential in continuous space. The tethering can be represented via a potential  $\mathcal{U}(n) = \frac{\kappa}{2}\,(n-R)^2$, with stiffness $\kappa>0$, which mimics a discrete harmonic trap in which the jump probability  decreases symmetrically as we approach the centre.  
The associated discrete restoring force $\mathfrak F(n)$ is obtained by taking a central-difference gradient, i.e., one has $\mathfrak F(n) \propto -(n-R)$, which pushes the walker back towards $R$ with a strength proportional to its distance from the centre.  
This is precisely the hallmark of an \emph{elastic} potential, whereby far from the centre the force is large, leading to a strong bias towards $R$, whereas the restoring force weakens near the centre.

For analytical convenience, as it will become apparent below, we consider the lattice walker within the elastic U--shaped potential to be confined between sites $n=0$ and site $n=2R+1$, with the minimum of the potential located at the central site $R$.  
The end sites $n=0$ and $n=2R$ act as reflecting boundaries, with the jump probabilities to remain still and to move inward, respectively, given by $1-q$, and $q$.

\begin{figure}[t]
\centering
\includegraphics[scale=1]{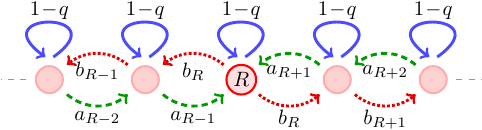}
\caption{Schematic diagram of transition probabilities for a walker under the U--shaped potential centred at the focal site $R$. The green (dashed) and  red (dotted) arrows denote space-dependent probabilities $a_n$ and $b_n$, respectively, given by $a_n = q(R+|n-R|)/(2R)$ and $b_n = q(R-|n-R|)/(2R)$. 
Note that one has $a_{R+n}=a_{R-n}$ and $b_{R+n}=b_{R-n}$, which make the potential symmetric about $R$.
The probability strength of the green (dashed) arrows weakens as we approach $R$, while that of the red (dotted) arrows weakens as we move away from $R$.
Imposing reflecting boundary conditions at sites 0 and $2R$ (which are not shown here) yields the Master equation~\eqref{eq:Master_Upot}.  }
\label{fig:upot-scheme}
\end{figure}

When in the bulk of the domain, at each time step the walker chooses its direction in a way that reflects the restoring force: the walker at site $m$ hops (i) to site $m-1$ (to the left) with probability $p_{m\to m-1} = q m / (2R)$ and  (ii) to site $m+1$ (to the right) with probability $p_{m\to m+1} = q [1- m / (2R)]$ (see Fig.~\ref{fig:upot-scheme}).
Thus, for $m<R$ we have $p_{m\to m+1} > p_{m\to m-1}$ and the walker drifts to the right, whereas for $m>R$ the inequality is reversed and the walker drifts to the left. 
The average displacement from site $m$ in a single time step is $\langle \Delta n \rangle_m = p_{m\to m+1} - p_{m\to m-1} = q (1-m/R)$, which changes sign at $m=R$, indicating a linear restoring drift towards the potential minimum.
In this way, the hopping rules implement a discrete drift that increases linearly with the displacement from the potential minimum, reproducing on the lattice the qualitative behaviour of the Ornstein--Uhlenbeck process in finite continuous space.

Let $P(n,t)$ denote the probability of finding the walker at site $n$ at time $t$;  
the corresponding Master equation reads~\cite{kac_random_1947}
\begin{align}
    P(0,t+1) &= (1-q) P(0,t) +  \frac{q}{2R}  P(1,t), \nonumber \\
    P(n,t+1) &= ( 1-q) P(n,t) +  q \qty(\frac{n+1}{2R}) P(n+1,t) \nonumber \\
    &\hskip-30pt + q \qty(1-\frac{n-1}{2R}) P(n-1,t), \quad n \in [1,2R-1],  \label{eq:Master_Upot} \\
    P(2R,t+1) &= (1-q) P(2R,t) +  \frac{q}{2R}  P(2R - 1,t) \nonumber.
\end{align}

\subsection{Propagator dynamics}

The propagator for Eq.~\eqref{eq:Master_Upot} is given by [see appendix~\ref{app:prop_upot}]
\begin{align}
    P(n,t|n_0) =\frac{1}{2^{2R}}
\sum_{j=0}^{2R} K_n(j)\,K_{j}(n_0) \,
\lambda_j^{t},
    \label{eq:prop_OU}
\end{align}
where $K_x(y)$ represents the Krawtchouk polynomials
\begin{align}
K_n(j) \equiv \sum_{m=0}^{n} (-1)^m \binom{j}{m}\binom{2R-j}{n-m},
\label{eq:Kkj-def}
\end{align}
with $0\le n,j\le 2R$, $\binom{j}{m}$ the binomial coefficient indexed by the integers $j \geq m \geq 0$, and
\begin{align}
    \lambda_j \equiv 1-j \frac{q}{R}.
    \label{eq:lambda_j-def}
\end{align}
In obtaining Eq. (\ref{eq:prop_OU}) we have generalised the method of solution employed in \cite{kac_random_1947} to solve Eq. (\ref{eq:Master_Upot}) when $q=1$ to arbitrary $q\leq 1$.

The propagator generating function 
\begin{align}
    \widetilde{P}(n,z|n_0) =  \frac{1}{2^{2R} } \sum_{j=0}^{2R}     \frac{K_n(j) K_j(n_0)}{1- z \lambda_j}  ,
\label{eq:prop_elastic_genZ}
\end{align}
obtained from Eq.~\eqref{eq:prop_OU},
allows to determine directly the steady-state via $P^{\mathrm{ss}}(n) \equiv \lim_{z\to 1} (1-z) \widetilde{P}(n,z|n_0) = \frac{1}{2^{2R} } \sum_{j=0}^{2R}   K_n(j) K_j(n_0) \delta_{j,0}$, which gives
\begin{align}
    P^{\mathrm{ss}}(n)  =  \frac{1}{2^{2R} } \binom{2R}{n}  \, ,
    \label{eq:ss-Upot}
\end{align}
where we have used Eq.~\eqref{eq:Kkj-def} with $K_0(n_0)=1$ and $\binom{0}{m} = \delta_{m,0}$. Similarly to the V--potential, the steady state in (\ref{eq:ss-Upot}) is independent of $q$,  and has the expected mean and variance of a binomial distribution, namely, $\expval{n}=R$ and ${\mathrm{Var}}(n) = R/2$.

\begin{figure}[t]
    \centering
    \includegraphics[scale=1]{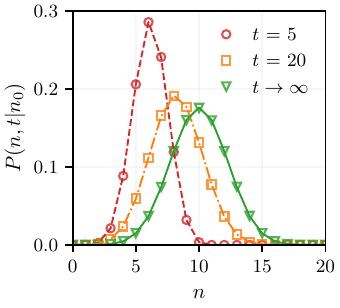}
    \caption{Propagator $P(n, t|n_0)$ for  dynamics~\eqref{eq:Master_Upot} under a U--potential centred at site $R=10$. The walker starts from site $n_0=5$ and we have $q=0.5$. The points are obtained from $5\times10^5$ stochastic realizations of dynamics~\eqref{eq:Master_Upot}. The broken lines are obtained from  Eq.~\eqref{eq:prop_OU}, while the continuous line in green denotes the steady-state given in Eq.~\eqref{eq:ss-Upot}.}
    \label{fig:upot_prop}
\end{figure}

The dynamics of the occupation probability is displayed in
Fig.~\ref{fig:upot_prop}, where $P(n,t|n_0)$ is plotted at different $t$ values. 
Compared to the dynamics for the V--shaped potential the U--shaped case exhibits a much smoother evolution towards the potential minimum  and the distribution retains a more  rounded shape during relaxation.

\subsection{First-passage statistics}

Employing Eq.~\eqref{eq:prop_elastic_genZ} it is straightforward to compute the first-passage generating function,
\begin{align}
    \widetilde{F}(n,z|n_0) =  \sum_{j=0}^{2R}     \frac{K_n(j) K_j(n_0)}{1- z \lambda_j} \qty[ \sum_{j=0}^{2R}     \frac{K_n(j) K_j(n)}{1- z \lambda_j}]^{-1}  ,
\label{eq:FPT_inZ_UPot}
\end{align}
and use the above expression to derive the mean first-passage time from $n_0$ to $n$ [see Appendix~\ref{sec:fpt-Upot-app}]
\begin{align}
\expval{T} &= \frac{R}{q \binom{2R}{n}}
\sum_{j=1}^{2R}
\frac{K_n(j)  K^{-}_{j}(n,n_0) }{j},
\label{eq:MFPT_Upot} 
\end{align}
and its corresponding variance 
\begin{align}
\mathrm{Var}(T) &= \expval{T} \qty[  \frac{R}{q\binom{2R}{n}} \sum_{j=1}^{2R} \frac{K_n(j)  K^{+}_{j}(n,n_0) }{j} - 1 ] \nonumber \\
    & +   \frac{2R^2}{q^2 \binom{2R}{n}} \sum_{j=1}^{2R} \frac{K_n(j)  K^{-}_{j}(n,n_0) }{j^2},
    \label{eq:VarFPT_Upot}
\end{align}
where $K^{\pm}_{j}(n,n_0) = K_j(n) \pm K_j(n_0)$.

\begin{figure}[t]
    \centering
    \includegraphics[scale=0.72]{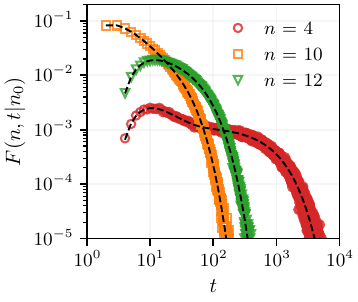}\hskip-2pt
    \includegraphics[scale=0.72]{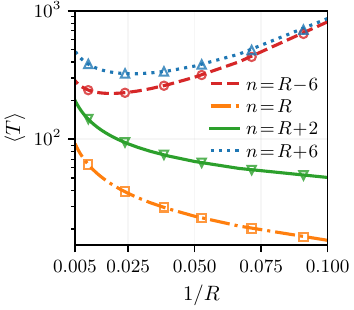}\\[0.2ex]
    \includegraphics[scale=0.72]{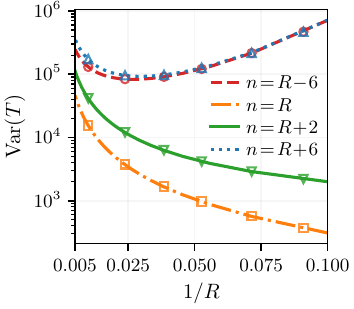}\hskip1pt
    \includegraphics[scale=0.72]{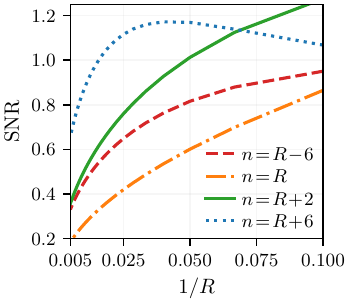}
    \caption{First-passage statistics with different target location $n$ for a U--potential centred at site $R$ and confined by reflecting boundaries at sites 0 and $2R$ with diffusivity $q=0.5$. In the plots, the points are obtained from $5\times10^5$ stochastic realizations of dynamics~\eqref{eq:Master_Upot}, while the lines denote analytical results. Top left: First-passage probability $F(n, t|n_0)$ as a function of time $t$ for $n_0=8$ and $R=10$ with black dashed lines obtained by numerically inverting Eq.~\eqref{eq:FPT_inZ_UPot}. 
    For the following plots $R$ is varied and the walker starts from $n_0 = R-2$.
    Top right: Mean first-passage time $\expval{T}$ from Eq.~\eqref{eq:MFPT_Upot}. Bottom left: Variance of the first-passage time $\mathrm{Var}(T)$ from Eq.~\eqref{eq:VarFPT_Upot}. Bottom right: The signal-to-noise ratio obtained using the relation $\mathrm{SNR}=\expval{T}^2/\mathrm{Var}(T)$.}
\label{fig:upot_fpt}
\end{figure}

The top--left panel of Fig.~\ref{fig:upot_fpt} shows the first--passage distribution
$F(n,t| n_{0})$ for the U--potential with $n_{0}=8$, 
The target at $n=10$, i.e., the bottom of the confining well, is reached more
rapidly, and the first-passage probability mode appears early on. When $n=12$, as the target lies uphill after the walker reaches the bottom of the potential, it produces a lower peak at later times with a
longer tail.  
For a target at $n=4$, the walker must move  constantly against the restoring force, making early arrivals unlikely. Most trajectories linger near the potential minimum before a rare excursion reaches the target, resulting in a noticeably broader and slower  decaying first-passage distribution than for $n=10$ or $n=12$.
Overall, the qualitative features of the FP dynamics closely mirror those observed for the
V--shaped potential in the top--left panel of Fig.~\ref{fig:vpot_fpt}.

The top--right panel of Fig.~\ref{fig:upot_fpt} shows the mean first–passage time $\langle T\rangle$ as a function of $1/R$, which controls the strength of the restoring force in the U–shaped potential.
In this case, increasing $1/R$ makes the potential narrower and steeper, thereby strengthening the pull towards the minimum at $n=R$, while decreasing $1/R$ corresponds to a broader and flatter confinement landscape with weaker restoring forces. The walker starts from $n_0=R-2$, slightly to the left of the minimum.
For the target located exactly at the minimum ($n=R$), the motion is always downhill so that strengthening the restoring force (i.e., increasing $1/R$) accelerates the descent and thereby causes the mean first–passage time $\langle T\rangle$ to decrease monotonically with $1/R$.
For the target at $n=R+2$, the walker first moves downhill to the minimum and then climbs only a short distance uphill. Because the restoring force is  relatively weak near $n=R$, this uphill motion is barely hampered, the main effect of increasing $1/R$ remains the same and $\langle T\rangle$ decreases monotonically also in this case. 
Targets farther from the minimum show a different trend. For $n = R-6$, the walker always has to go against the restoring force. When $1/R$ is small (a very broad and shallow potential), the walker can roam far away before eventually reaching the target, so $\langle T\rangle$ is large.
As $1/R$ increases, the restoring force limits these long excursions and $\langle T\rangle$ decreases. However, if $1/R$ becomes too large, the potential becomes steep and moving left becomes very unlikely. The walker then reaches $n=R-6$ only through rare uphill events, so $\langle T\rangle$ starts to grow again.
For $n = R+6$, the walker first goes downhill to the minimum and then must climb a long uphill stretch. A moderate restoring force keeps the walker near the centre and speeds up arrival, but when the force becomes strong the uphill climb dominates. Reaching the target then requires a rare excursion far to the right, leading again to an increase in $\langle T\rangle$ at large $1/R$.
These qualitative features of the mean are also mimicked by the variance, plotted in the bottom--left panel.

The SNR for the U--potential is depicted in the bottom--right panel of Fig.~\ref{fig:upot_fpt}.
As we have discussed, for targets near the potential minimum ($n=R$ and $n=R+2$), the restoring force either fully assists the motion or only weakly opposes it over a very short distance.
With increasing $1/R$, as the fluctuations decrease faster than the mean in this case, the SNR increases monotonically for both cases.
For a far target on the left ($n=R-6$), the restoring force always pushes the walker in the opposite direction.
A stronger confinement makes leftward motion rarer and more variable, so the SNR stays close to~1. 
For the far–right target ($n=R+6$), weak confinement improves predictability, but strong confinement makes the required uphill excursion increasingly rare, which enhances fluctuations and causes the SNR to peak at intermediate $1/R$.

\begin{figure}[t]
    \centering
    \includegraphics[scale=1]{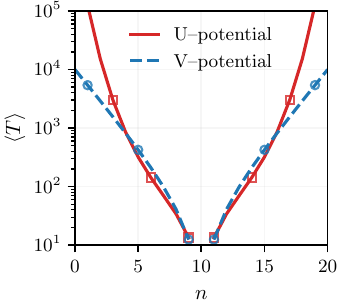}
    \caption{Comparison between the mean first-passage time $\expval{T}$ as function of target position $n$ for the V--shaped and U--shaped potentials, both centred at $R=10$ and limited by reflecting boundaries at 0 and $2R$. In both cases, the walker starts from the centre of the confinement, i.e.,  $n_0 =R=10$. Here, we have $q=0.5$ and a $g\simeq 0.295$ obtained by imposing the condition of equal variance at steady state in both cases, namely the right-hand side of Eq.~\eqref{eq:vpot-fully-var-n} for the V--potential and $R/2$ for the U--potential. The blue dashed line and the red continuous line are obtained using Eqs.~\eqref{eq:vpot-fully-mfpt} and~\eqref{eq:MFPT_Upot}, respectively. The points are obtained using $5\times 10^5$ stochastic realizations for both dynamics. The plot does not show any data point at the potential centre $n=10$, as the walkers start from the bottom of the potential.}
    \label{fig:compare-mfpt}
\end{figure}
Figure~\ref{fig:compare-mfpt} compares the mean first--passage time $\langle T \rangle$ as a function of the target position $n$ in the reflecting domain $[0,2R]$ for both V--shaped and U--shaped potentials centred at $n=R=10$. In order to get a meaningful comparison, we choose the value of $g$ for the V--potential so that the variances of the walker's position in the steady state become equal.
While both curves display a non-linear increase as the distance from $R$ increases,  their qualitative difference  can be understood from the nature of the restoring forces in the two confinements.  
Near the centre $n=R$, the V--shaped potential creates a \emph{constant} drift towards $R$, so the walker is  pulled sharply back whenever it moves away from the centre. 
As a result, moving against the drift is unlikely, which makes $\langle T\rangle$ larger than that in the U--shaped case, where the restoring force is  much smaller near $R$ and grows only gradually with distance.
In contrast, far from the centre, the U--shaped potential exerts a  \emph{stronger} restoring force than the V--shaped one because its effective drift increases with the distance from $R$, leading to much larger first--passage times.

\section{Confined walks with resetting}
\label{sec:reset}

To understand further how robust is the deterministic drift towards the bottom of the potential, we superimpose a stochastic resetting dynamics to the lattice walker, while moving under the V-- and U--potentials.  
The dynamics is as follows. 
At each discrete time step, the walker either (i) resets to a fixed site $n_r$ with probability $r$, or (ii) performs its underlying reset--free Markov step with probability $(1-r)$.
Let $W_{0}(n,t|n_0)$ and $\widetilde W_{0}(n,z|n_0)$ denote the propagator and its generating function, respectively, of the reset--free walk.
The propagator generating function in the presence of resetting is given by~\cite{das_discrete_2022}
\begin{align}
    \widetilde {\mathcal W}_r (n, z|n_0) &= \frac{z r}{1-z} \widetilde W_0 (n, (1-r)z|n_r) \nonumber \\ 
    &+ \widetilde W_0 (n, (1-r)z|n_0),
\label{eq:prob_with_reset}
\end{align}
with a steady-state given by~\cite{das_discrete_2022}
\begin{align}
    \mathcal W^{\mathrm{ss}}_r(n) = r \widetilde W_0 (n, 1-r|n_r) .
    \label{eq:reset-ss}
\end{align}

To study the LRW dynamics under the V and U--potentials in the presence of resetting in bounded domain $[0,2R]$, one simply needs to replace the reset-free propagator $\widetilde W_0(n,z|n_0)$ appearing in Eqs.~\eqref{eq:prob_with_reset} and~\eqref{eq:reset-ss}  with $ \widetilde Q^{\mathrm{ref}}(n,z | n_0) $ (given in Eq.~\eqref{eq:Vpot-fully-bound-prop}) and $\widetilde{P}(n,z|n_0)$ (given in Eq.~\eqref{eq:prop_elastic_genZ}), respectively. 
Hence, the steady state for the V--potential in the presence of resetting is
\begin{align}
    \mathcal Q^{\mathrm{ss}}_r(n) 
    &= r \widetilde L(n, 1-r | n_r) + r    \widetilde  L(2R, 1-r | n_r) \nonumber \\
    &\hskip-30pt \times \frac{
     \rho_r
    \big[ \widetilde L(n,1-r | 2R) - \widetilde L(n,1-r | 2R+1) \big] }
    {2 - \rho_r
      \big[ \widetilde L(2R,1-r | 2R) - \widetilde L(2R,1-r | 2R+1) \big] },
\label{eq:reset-v-ss}
\end{align}
where we have $\rho_r \equiv q (1-r) (1-g)$, while the steady state for the U--potential in the presence of resetting is
\begin{align}
    \mathcal P^{\mathrm{ss}}_r(n) =  \frac{r}{2^{2R} } \sum_{j=0}^{2R}     \frac{K_n(j) K_j(n_r)}{1- (1-r) \lambda_j}   .
    \label{eq:reset_U_ss}
\end{align}

\begin{figure}[t]
    \centering
    \includegraphics[scale=0.73]{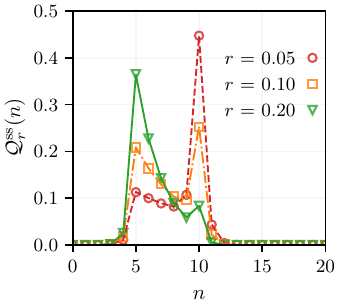} 
    \includegraphics[scale=0.73]{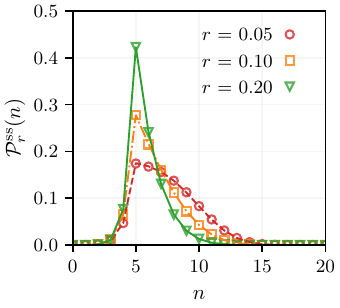}
    \caption{Steady-state probability distributions in the presence of resetting within the reflecting domain $[0,2R]$ for different values of reset probability $r$, with the potentials centred at $R=10$ and resetting site $n_r=5$ and $q=0.5$. Left: steady state $\mathcal Q^{\mathrm{ss}}_r(n)$ for the V--potential with $g=0.8$. Lines denote the analytical result~\eqref{eq:reset-v-ss}, while points are obtained from $5\times10^5$ stochastic realizations of dynamics~\eqref{eq:deb_vpot_master_compact} with resetting and reflecting boundaries at $0$ and $2R$. Right: steady state $\mathcal P^{\mathrm{ss}}_r(n)$ for the U--potential. Lines are given by the analytical result~\eqref{eq:reset_U_ss}, and points are obtained from $5\times10^5$ stochastic realizations of dynamics~\eqref{eq:Master_Upot} with resetting.}
    \label{fig:reset-ss}
\end{figure}
Figure~\ref{fig:reset-ss} shows the steady-state probability for a walker subject to stochastic resetting to the site $n_r=5$ evolving under the V--shaped (left panel) and U--shaped (right panel) potential within the reflecting domain $[0,2R]$ and the minimum at $n_c=R$.  The stationary distribution results from a competition between confinement by the potential and repeated returns to the reset site, leading to a nonequilibrium steady state whose shape depends on the reset probability $r$ in markedly different ways in the two cases. For the V--potential, while the steady state is primarily controlled by the confining potential, a smaller peak appears around $n_r$ even for small $r$.
The same cannot be said for the U--potential, where  the confining force does not induce a  strong localization, but  skews the probability towards $n_c$,  resulting in a single broadened peak around $n_r$, whose position and width are controlled by the balance between the elastic potential and resetting.
Similar steady-state behaviour of the occupation probability for diffusion under V-- and U--potentials in continuous case is reported in Ref.~\cite{pal_diffusion_2015}.

Closed analytical expressions for the mean position and its variance in the steady state of the confined V--potential are  rather long as they employ Eq.~\eqref{eq:reset-v-ss} with Eqs.~\eqref{eq:Vpot-left-bound-prop} and~\eqref{eq:GF_inZ_VPot_gen_sol}. 
In contrast, for the U--potential these quantities can be obtained straightforwardly by exploiting the generating function of $K_n(j)$ given in Eq.~\eqref{eq:Knj-GF}, which yields $\sum_{n=0}^{2R} n K_n(j) = 2^{2R-1}(2R\delta_{j,0}-\delta_{j,1})$ and $\sum_{n=0}^{2R} n^2 K_n(j) = 2^{2R-1}[R(2R+1) \delta_{j,0} -2R \delta_{j,1} +\delta_{j,2} ]$.
Using these relations and Eq.~\eqref{eq:reset_U_ss}, one can explicitly compute the mean and variance of the position in the steady state of the U--potential with resetting, leading finally to
\begin{align}
    \expval*{n} &= \frac{ (r n_r +q -q r) R}{r R + q -q r} , \\
    \mathrm{Var}(n) &= \frac{q (1-r) R }{(2 q (1-r)+r R) (r R+q-qr)^2} \nonumber \\
    &\hskip-27pt \times [q r (1-r) \left((n_r-R)^2+2 R\right)+q^2 (1-r)^2+r^2 R^2 ] .
    \label{eq:reset-u-ss-var}
\end{align}
As anticipated, both the mean position  and the variance now show an explicit dependence on the resetting site $n_r$, and additionally they depend on the diffusivity $q$, making it clear that resetting fundamentally alters the nature of the steady state. 

\subsection{First-passage statistics}

\begin{figure}[t]
    \centering
    \includegraphics[scale=0.73]{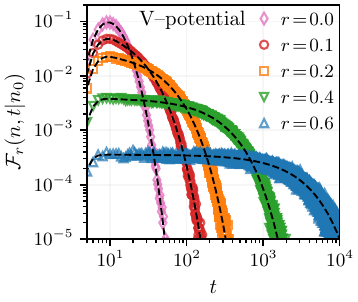} 
    \includegraphics[scale=0.73]{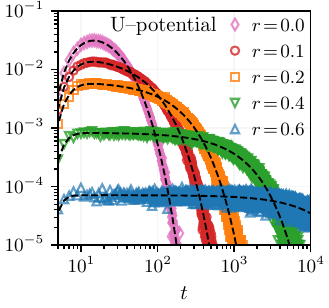}
    \caption{First-passage probability starting from the resetting site $(n_0=n_r=5)$ to the centre of the potential $(n=R=10)$ in reflecting domain $[0,2R]$ for different values of reset probability $r$ and $q=0.5$. The left and right panels depict, respectively, the dynamics in the V--potential ($g=0.8$) and the U--potential. Points are obtained from $5\times10^5$ stochastic realizations, while black dashed lines are computed by numerically inverting Eq.~\eqref{eq:reset-fpt} by replacing $\widetilde{W}_{0}$ with $ \widetilde Q^{\mathrm{ref}}(n,z | n_0) $ and $\widetilde{P}(n,z|n_0)$ for the V-- and U--potential, respectively.}
    \label{fig:reset-fpt}
\end{figure}

To study the dynamics of the first-passage probability we take its generating function given by~\cite{das_discrete_2022}
\begin{align}
    &\widetilde{\mathcal{F}}_r (n,z|n_{0}) = \nonumber \\
    &\frac{ r z \widetilde{W}_{0}(n,(1-r)z|n_{r})
    +(1-z)\,\widetilde{W}_{0}(n,(1-r)z|n_{0})}{
    r z \widetilde{W}_{0}(n,(1-r)z|n_{r})
    +(1-z)\widetilde{W}_{0}(n,(1-r)z|n) } ,
\label{eq:reset-fpt}
\end{align}
and invert it numerically~\cite{abate_numerical_1992}.
Figure~\ref{fig:reset-fpt} shows the first-passage probability $\mathcal{F}_r(n,t| n_0)$ to the potential centre $n=R$, starting from the resetting site $n_0=n_r$, for different reset probabilities $r$. For both the V-- and U--potentials, increasing $r$ enhances localization near the reset site $n_r$, thereby reducing the likelihood of excursions that reach the target at the centre. As a result, first-passage events become increasingly rare and are pushed to longer times, leading to broader distributions with more pronounced long-time tails. The qualitative behaviour is common to both potentials, while quantitatively the V--potential exhibits slightly sharper short-time features due to its constant drift towards the centre. For both potentials larger values of $r$ make the tail of the distribution flatter, signalling the onset of the motion limited regime. 
The broadening of $\mathcal{F}_r(n,t| n_0)$ is also manifest (not shown) in the mean first-passage time $\expval{\mathcal T_r}$ to reach $n$  from $n_0$ that can be readily extracted from the propagator expression via~\cite{das_discrete_2022}
\begin{equation}
\expval{\mathcal T_r}
 = \frac{\widetilde{W}_{0}(n,1-r|n) - \widetilde{W}_{0}(n,1-r|n_0)}{r\,\widetilde{W}_{0}(n,1-r|n_r)} .
\label{eq:reset-mfpt}
\end{equation}

\begin{figure}[t]
    \centering
    \includegraphics[scale=0.74]{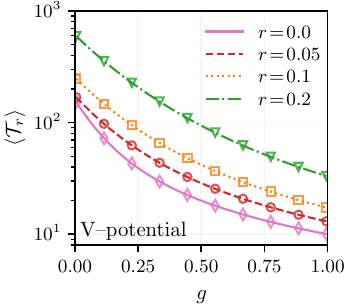} 
    \includegraphics[scale=0.74]{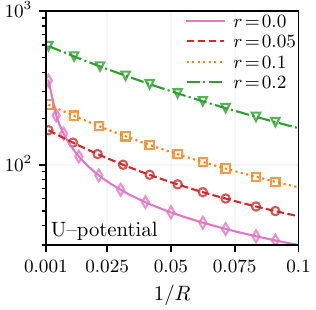}
    \caption{Mean first-passage time $\expval{\mathcal{T}_r}$ to the potential minimum $(n=R)$ starting from the resetting site $(n_0=n_r)$ in reflecting domain $[0,2R]$ with $q=0.5$ for different values of reset probability $r$. The left panel depicts $\expval{\mathcal{T}_r}$ for the V--potential with $n_r=5$ and $R=10$. The right panel shows the same for the U--potential with $n_r =R-5$. Points are obtained from $5\times10^5$ stochastic realizations, while the lines are computed using Eq.~\eqref{eq:reset-mfpt} by replacing $\widetilde{W}_{0}$ with $ \widetilde Q^{\mathrm{ref}}(n,z | n_0) $ and $\widetilde{P}(n,z|n_0)$ for the V-- and U--potential, respectively.}
    \label{fig:reset-mfpt}
\end{figure}

Figure~\ref{fig:reset-mfpt} illustrates the interplay between confinement and resetting on the mean first-passage time to the potential centre starting from the resetting site. 
For both V-- and U--potentials, increasing the bias strength ($g$ for V--potential and $1/R$ for U--potential) reduces $\langle \mathcal{T}_r \rangle$ for all resetting probabilities due to enhanced drift towards the target, while for fixed bias, increasing $r$ prolongs the first-passage time by interrupting this directed motion. 
For the U--potential with $r=0$, $\langle \mathcal{T}_r \rangle$ grows rapidly with $R$ because increasing the system size weakens the restoring force near the centre, so that in the large--$R$ limit the dynamics close to the potential minimum becomes nearly unbiased, leading to a pronounced increase in the first-passage time.
This behaviour suggests that, in contrast to the V--potential where a constant bias towards the centre persists throughout the lattice, resetting can become beneficial in the U--potential for sufficiently large $R$, in the sense that it may reduce the mean first-passage time relative to the no-reset case. 
\subsection{Average number of distinct sites visited}

Similar to Eq.~\eqref{eq:MNDSV}, the generating function of the  average number of distinct visited sites in the presence of resetting simply reads~\cite{biroli_number_2022}
\begin{align}
\expval*{\widetilde{\mathcal N}_r(z) } = \frac{1}{1-z}\sum_{n} \widetilde{\mathcal{F}}_r(n,z| n_0) ,
\label{eq:navg-in-z}
\end{align}
with $\widetilde{\mathcal F}_r(n,z|n_0)$ defined in Eq.~\eqref{eq:reset-fpt}.
In spatially bounded reflecting domain $[0,2R]$, irrespective of the confining potential the mean $\expval{\mathcal N_r(t)}$ saturates  to the domain size $2R+1$ as $t\to\infty$, i.e., the walker eventually covers all accessible space.

\begin{figure}[t]
    \centering
    \includegraphics[scale=0.78]{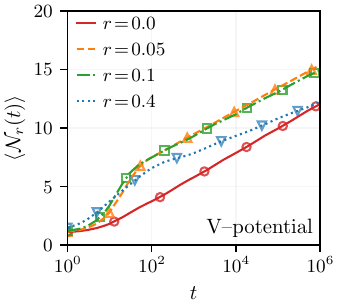}
    \includegraphics[scale=0.78]{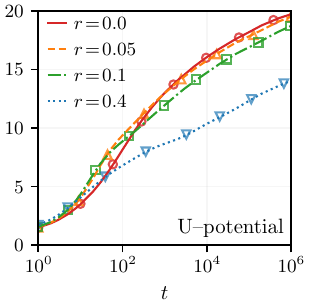}
    \caption{Average number of distinct sites visited by a walker as a function of time $t$ for different values of reset probability $r$. The walker starts from the centre of the potential $(n_0=R=10)$ in the reflecting domain $[0,2R]$ with $q=0.5$ and the resetting site at $n_r=5$. Left: V--potential with $g=0.8$. Right: U--potential. Points are obtained from $5\times10^5$ stochastic realizations, while lines are computed by numerically inverting Eq.~\eqref{eq:navg-in-z} along with Eq.~\eqref{eq:reset-fpt} by replacing $\widetilde{W}_{0}$ with $ \widetilde Q^{\mathrm{ref}}(n,z | n_0) $ and $\widetilde{P}(n,z|n_0)$ for the V and U--potential, respectively.}
    \label{fig:reset-dsv}
\end{figure}

Figure~\ref{fig:reset-dsv} shows the time evolution of $\langle \mathcal N_r(t)\rangle$, for the walker evolving under the V (left panel) and U (right panel) potentials starting from the focal point. 
In the absence of resetting ($r=0$), both cases display a monotonic growth of $\langle \mathcal N_r(t)\rangle$ as the walker progressively explores the finite domain; however, the rate of growth is faster for the U--potential than for the V--potential. 
This difference originates from the stronger drift towards the centre in the V--potential, which repeatedly pulls the walker back and limits its ability to visit new sites, whereas the smoother U--potential allows for broader exploration as the restoring force is negligible in the neighbourhood of the centre.  
When resetting is introduced ($r>0$), the exploration dynamics changes qualitatively in both potentials. At short times, even for a tiny non-zero $r$, $\langle \mathcal N_r(t)\rangle$ initially grows more rapidly than in the reset-free case, because only a single resetting event to  $n_r$ is enough to produce a long-range jump and hence promote exploration of new sites in the neighbourhood of $n_r$. 
However, large values of $r$ increasingly restricts exploration: frequent resets confine the walker near $n_r$, reducing the rate at which new sites are discovered and leading to a clear suppression of $\langle \mathcal N_r(t)\rangle$ compared to the $r=0$ case.

\section{Discussions}
\label{sec:discus}

We have studied the dynamics of a diffusing lattice random walker subject to the restoring force of two types of focal point potential, a V--shaped and a U--shaped one. For the former, we have been able to solve exactly the Master equation that describes the dynamics of the occupation probability in unbounded space. Using the Master equation propagator, we have analysed the first-passage dynamics to a single target and shown that, depending on the relative locations of target and initial site, there exists an optimal mean first-passage time as a function of the bias strength. By exploring the dynamics of the average number of distinct sites visited, we have identified a long-time logarithmic growth and found its scaling coefficient. Employing the defect technique, we have also found analytically the generating function solution of the semi-bounded and fully bounded propagator, as well as the first passage probability and its mean in the latter case.

For the U--potential, we have considered a fully bounded domain and  found exactly the time-dependent propagator solution of the Master equation in terms of the (orthogonal) Krawtchouk polynomials. In comparing the first-passage dynamics with the fully bounded V--potential, we find that the two potentials display the same qualitative features, including the existence of a minimum in the mean first-passage time as a function of the bias for the V--potential and the domain size for the U--potential.

The inclusion of resetting dynamics to the lattice walker in the two potentials has unearthed interesting effects. For the V--potential, the occupation probability displays a double peak at steady state, one at the resetting site and one at the potential minimum. On the other hand, using the same resetting probability and the same domain size as for the V--potential, the occupation probability for the U--potential case does not display a double peak, but skews its shape towards the potential minimum.

The strong dependence of the resetting dynamics even in the presence of a confining potential becomes evident when looking at the first-passage statistics. We have, in fact, uncovered that even for intermediate values of the resetting probability, the time dependence of the first-passage probability exhibits features (flat distribution for very long times) characteristic of the motion-limited regime.

Overall, our analysis on diffusion within a focal point potential represents the introduction of new analytical tools to quantify the spatio-temporal dynamics of LRWs in a heterogeneous environment. Given the advantage of a discrete representation, compared to one using continuous variables, when studying diffusion in a disordered space, our findings provide the means to extend analytic predictions to when deterministic biases and random disordered are present simultaneously, a feature naturally present in realistic systems.

\section*{Acknowledgments}
DD acknowledges support from the MUR PRIN2022 project ``Breakdown of ergodicity in classical and quantum many-body systems'' (BECQuMB) Grant No. 20222BHC9Z.
LG acknowledges funding from the National and Environmental Research Council (NERC) Grant No. NE/W00545X/1. 

\section*{Data availability}
All data that support the findings of this study are included within the article.

\appendix

\begin{widetext}

\section{Unbounded propagator in V--shaped confinement}
\label{sec:Vpot-prop-app}

\subsection{Propagator generating function}

We consider the lattice walk dynamics in heterogeneous space with two media separated by the Type B interface placed at site $M$, such that sites $n \leq (M-1) $ and $n \geq (M+1)$ belong to medium--1, with diffusivity $q_1$ and bias $g_1$, and  medium--2, with diffusivity $q_2$ and bias $g_2$, respectively~\cite{das_dynamics_2023}.
The interface at the shared site $M$ is associated with a hypothetical third medium--$c$.
Denoting the probability to find the walker
at site $n$ in medium--$\mu$ at time $t$ by $P_\mu(n, t)$, one may write the Master equation:
\begin{align}
& P_1 (n,t+1) = (1-q_1) P_1(n,t) + \frac{q_1}{2} (1-g_1)  P_1(n-1,t) + \frac{q_1}{2} (1+g_1) P_1(n+1,t)   \, ; \quad n < (M-1)  \, , \nonumber \\
& P_1 (M-1,t+1) = (1-q_1) P_1(M-1,t) + \frac{q_1}{2} (1-g_1)  P_1(M-2,t) + \frac{q_1}{2} (1+g_1) P_c(M,t)   \, , \nonumber \\
& P_{c} (M,t+1) = \left[ 1 - \frac{q_1}{2} (1+g_1) - \frac{q_2}{2} (1-g_2) \right] P_c (M,t) + \frac{q_1}{2} (1-g_1) P_1(M-1,t) + \frac{q_2}{2} (1+g_2) P_2(M+1,t)    \, , \label{eq:master_eq_3_V}\\
& P_2 (M+1,t+1) = (1-q_2) P_2(M+1,t) + \frac{q_2}{2} (1-g_2) P_c(M,t) + \frac{q_2}{2} (1+g_2) P_2(M+2,t) \, , \nonumber \\
& P_2 (n,t+1) = (1-q_2) P_2(n,t) + \frac{q_2}{2} (1-g_2) P_2(n-1,t) + \frac{q_2}{2} (1+g_2) P_2(n+1,t) \, ; \quad n > (M+1) \, . \nonumber  
\end{align}

The coupled system of Eqs.~(\ref{eq:master_eq_3_V})
can be solved exactly in terms of the propagator generating function
$S_{\mu,\nu}(n,z|n_0)$, where the indices $\mu$ and $\nu$ denote the medium
containing the observation site $n$ and the initial site $n_0$,
respectively (with $\mu,\nu \in \{1,2,c\}$).
Following Ref.~\cite{das_dynamics_2023}, the solution proceeds by
introducing discrete Fourier transforms of the unknown functions
$G_{1,\nu}(u,z|n_0)=\sum_{n=-\infty}^{M-1} u^n\,S_{1,\nu}(n,z|n_0)$ and 
$G_{2,\nu}(u,z|n_0)=\sum_{n=M+1}^{\infty} u^n\,S_{2,\nu}(n,z|n_0)$,
which, together with the central-site component $S_{c,\nu}(n,z|n_0)$,
lead to a closed algebraic system of three linear equations.
Solving this system yields explicit expressions for
$S_{1,\nu}$, $S_{2,\nu}$, and $S_{c,\nu}$ for each of the three possible
initial condition sectors $\nu \in \{1,2,c\}$.
The complete propagator is made up of nine distinct
generating functions $\{S_{\mu,\nu}\}$, based on all possible initial and final regions~\cite{das_dynamics_2023}:
\begin{align}
S_{1,1}(n, z|n_0) &= \frac{1}{\sqrt{1-\beta^{+}_1  \beta^{-}_1}} \Bigg[ \frac{ f_1^{ \frac{ n - n_0 + |n-n_0|}{2} } \xi_1^{|n-n_0|} }{1-z + z q_1} - f_1^{M-n_0}  \xi_1^{2M-n-n_0} \bigg\{ \frac{1}{1-z+ z q_1} - \frac{1}{\Gamma}  \sqrt{1-\beta^{+}_1  \beta^{-}_1} \bigg\} \Bigg]   , \nonumber \\
S_{2,1}(n, z|n_0) &= \frac{1}{\Gamma} \, (f_1 \xi_1)^{M-n_0} \,   (f_2\xi_2)^{n-M}   , \qquad 
S_{c,1}(n,z|n_0)  = \frac{1}{\Gamma}  (f_1\xi_1)^{M-n_0} , \qquad
S_{1,2}(n,z|n_0) = \frac{1}{\Gamma}  \xi_1^{M-n}  \xi_2^{n_0-M}  ,  \nonumber \\[-1ex]
\label{eq:app-tB-9sols} \\[-1ex]
S_{2,2}(n, z|n_0) &= \frac{1}{\sqrt{1-\beta^{+}_2  \beta^{-}_2}} \Bigg[ \frac{ f_2^{ \frac{ n - n_0 + |n-n_0|}{2} } \xi_2^{|n-n_0|} }{1-z+ z q_2} - f_2^{n-M} \, \xi_2^{n+n_0-2M} \bigg\{ \frac{1}{1-z+ z q_2} - \frac{1}{\Gamma} \, \sqrt{1-\beta^{+}_2  \beta^{-}_2} \bigg\} \Bigg]  , \nonumber \\
S_{c,2}(n,z|n_0)  &= \frac{1}{\Gamma}  \xi_2^{n_0-M}  ,        \qquad
S_{1,c}(n, z|n_0) = \frac{1}{\Gamma}  \xi_1^{M-n} ,            \qquad   
S_{2,c}(n, z|n_0) = \frac{1}{\Gamma}  (f_2\xi_2)^{n-M} , \qquad 
S_{c,c}(n,z|n_0) = \frac{1}{\Gamma}  , \nonumber 
\end{align}
where
\begin{align}
& \beta^{\pm}_\mu(z) \equiv  \frac{z q_\mu (1\pm g_\mu)}{1-z(1-q_\mu)} , \qquad f_\mu \equiv \frac{1-g_\mu}{1+g_\mu} = \frac{\beta^{-}_\mu}{\beta^{+}_\mu} , \qquad \xi_\mu(z)  \equiv  \frac{1-\sqrt{1-\beta^{+}_\mu(z) \beta^{-}_\mu(z)}}{\beta^{-}_\mu(z)}  \nonumber  \\[-1ex] \label{eq:betaPM_i_def} \\[-1ex]
& \Gamma \equiv 1-z + \frac{z q_1 }{2} \Big[ 1  + g_1 - \xi_1 (1- g_1) \Big]  + \frac{z q_2 }{2} \Big[1 - g_2 - f_2 \,  \xi_2 (1+g_2) \Big] . \nonumber 
\end{align}
Accounting for all nine combinations of $\mu$ and $\nu$, the general solution of the Master equation~(\ref{eq:master_eq_3_V}) may be written in terms of the propagator generating function using the discrete Heaviside function $\Theta(n)$ as~\cite{das_dynamics_2023}
\begin{align}
S(n,z|n_0) &= \Theta(M-1-n_0) \Big[ \Theta(M-1-n) S_{1,1}(n,z|n_0) + \delta_{n,M}\,  S_{c,1}(n,z|n_0) + \Theta(n-M-1)  S_{2,1}(n,z|n_0) \Big] \nonumber \\
& +\delta_{n_0,M}   \Big[ \Theta(M-1-n) S_{1,c}(n,z|n_0) + \delta_{n,M}\,  S_{c,c}(n,z|n_0) + \Theta(n-M-1)  S_{2,c}(n,z|n_0) \Big] \nonumber \\
& + \Theta(n_0-M-1) \Big[ \Theta(M-1-n) S_{1,2}(n,z|n_0) + \delta_{n,M}\,  S_{c,2}(n,z|n_0) + \Theta(n-M-1)  S_{2,2}(n,z|n_0) \Big]  \, . \label{eq:GF_inZ_tB_gen_sol}
\end{align}

The lattice walk dynamics~\eqref{eq:deb_vpot_master_compact} under the V--potential can be mapped exactly to  dynamics~\eqref{eq:master_eq_3_V}  with $q_1=q_2=q$, $g_1 = -g_2 = - g;~g>0$, and the focal point set at the Type B interface, i.e., $M=n_c$. This exact mapping yields from Eq.~\eqref{eq:betaPM_i_def} a set of  helpful relations: 
\begin{align}
& \beta^{+}_2 = \beta^{-}_1  = \frac{z q(1+g)}{1-z(1-q)} \equiv \beta_+ \, , \qquad  
\beta^{-}_2 = \beta^{+}_1  = \frac{z q(1-g)}{1-z(1-q)} \equiv \beta_-  , \qquad 
\beta^{+}_1  \beta^{-}_1 = \beta^{+}_2  \beta^{-}_2 = \beta_{+} \beta_{-}, \nonumber \\
& f_2 = f_1^{-1} = \frac{1-g}{1+g} \equiv f \, , \qquad  \xi_1  = f_2  \xi_2 = f \xi_2  =  \frac{1-\sqrt{1-\beta_{+} \beta_{-}}}{\beta_{+}} \equiv \xi , \qquad  \Gamma  \equiv \chi .
\end{align}
Using the above definitions of $\chi$, $f$, $\beta_{\pm}$, and $\xi$, which are defined as Eqs.~\eqref{eq:chi-def} and~\eqref{eq:xi-def} of the main text, one can significantly reduce the complexity of the functions $S_{\mu,\nu}(n,z|n_0)$ in Eq.~\eqref{eq:app-tB-9sols}.
Using $\alpha \equiv |n_c-n|+|n_c-n_0|-|n-n_0|$, it is possible to write a single expression for the general solution $S(n,z|n_0)$ in Eq.~\eqref{eq:GF_inZ_tB_gen_sol} which has been presented  in the main text as $Q(n,z|n_0)$ in Eq.~\eqref{eq:GF_inZ_VPot_gen_sol}.

\subsection{The steady-state}

We note that, at $z=1$, we have 
$\beta_\pm(1)  = 1\pm g$, $\xi(1) = f$, and $\chi(1)=0$. Therefore, in the limit $z \to 1$, only the single factor $\chi(z)$ generates a simple pole at $z=1$, while all other pieces remain finite. Expanding the propagator $\widetilde Q(n, z | n_0)$ from Eq.~\eqref{eq:GF_inZ_VPot_gen_sol} into a series of $(1-z)$, we obtain the leading behaviour as
\begin{align}
\widetilde Q(n, z | n_0) \approx \frac{{Q_0}}{1-z}+{Q_1}+{Q_2} (1-z)+{Q_3} (1-z)^2 + \mathcal O(1-z)^3,
\label{eq:Qtilde-pole}
\end{align}
where the $z$--independent coefficients $Q_j$ may depend on $n$, $n_0$, and are given by
\begin{align}
    Q_0 &\equiv g f^{|n-n_c|} ,\\
    Q_1 &\equiv \frac{2 f^{-\alpha/2} -g^2 (1-2 q) -2 Y -1}{2 g q}, \\
    Q_2 &\equiv \frac{4 f^{-\alpha/2} \left(g^2 q -g | n-n_0| -1\right) +2 Y \left(Y+g^2 (1-4 q)+2\right)+g^4 (1-2 q)^2-4 g^2 q+3}{4 g^3 q^2} , \\
    Q_3 &\equiv \frac{1}{24 g^5 q^3} \Big[  12 f^{-\alpha /2} \left(g | n-n_0|  \left(g | n-n_0| -4 g^2 q+3\right)+2 g^4 q^2-g^2 (4 q+1)+3\right)   \nonumber \\
    &\hskip45pt - 2 Y \left\{ Y \left(2 Y+g^2 (3-18 q)+9\right)+3 g^4 ( 1-6 q + 12 q^2 )+g^2 (1-36 q)+18 \right\} \nonumber \\
    &\hskip45pt - 3 g^6 (1-2 q)^3-36 g^4 q^2+9 g^2 (6 q+1)-30  \Big]   ,
\end{align}
with 
\begin{align}
    Y \equiv g (| n-n_c| +| n_0-n_c| ) .
\end{align}
The steady-state is easily obtained from Eq.~\eqref{eq:Qtilde-pole} using the final value theorem, which gives   $Q^{\rm ss}(n) = \lim_{z\to 1}(1-z)\,\widetilde Q(n,z| n_0)$, which produces Eq.~\eqref{eq:VPot_ss} of the main text.

\subsection{Time-dependent propagator}

We now provide a recipe to find the time-dependent propagator for the V--potential in unbounded space. To this end, one needs to apply a $z$-inversion to the generating function $\widetilde{Q} (n,z|n_0)$ given in  Eq.~\eqref{eq:GF_inZ_VPot_gen_sol}.
For later convenience, we first identify from Eqs.~\eqref{eq:chi-def} and~\eqref{eq:xi-def} the following relations
\begin{align}
\xi(z) \!=\! \frac{1-z(1-q)\!-\!\sqrt{( 1 \!-\! z \rho_+ \!) (1\!-\!z \rho_- \!)}}{zq(1+g)}, \quad
{\chi(z)+z q g} \!=\! {\sqrt{( 1\!-\!z \rho_+ \!) (1\!-\!z \rho_- \!)}}, \quad
\frac{1}{\chi(z)} \!=\! \frac{zqg+ \chi + z q g}{ (1\!-\!z)\big[1\!-\!z(1-2q)\big]} ,
\label{eq:app-chi-xi-chi-1}
\end{align}
where $\rho_\pm = 1-q \pm q\sqrt{1-g^2}$ and  rewrite $\widetilde{Q} (n,z|n_0)$ from Eq.~\eqref{eq:GF_inZ_VPot_gen_sol} as 
\begin{align}
    \widetilde Q(n,z|n_0) 
&= f^{-|n_c-n_0|} \xi^{|n-n_0|} \qty[ \frac{f^{\alpha/2}   }{  \chi + z q g } - \frac{ \xi^{\alpha} }{  \chi + z q g } +  \frac{q g z \xi^{\alpha} }{(1-z) \qty[1-z (1-2 q)]}  
+ \frac{ (\chi+zq g)\xi^{\alpha} }{(1-z) \qty[1-z (1-2 q)]}     ] \, .
\label{eq:S_z}
\end{align}
It is evident from Eq.~\eqref{eq:S_z} that to obtain the time-dependent propagator $Q(n,t|n_0)$ one needs to invert $z$-dependent functions of the following form
\begin{align}
    \widetilde F_1 (z,\gamma) \equiv \frac{ \xi^\gamma}{\chi +  z q g } , \quad 
    \widetilde F_2(z, \gamma) \equiv \frac{ z \xi^\gamma}{(1-z) \qty[1-z(1-2 q)]} , \quad 
    \widetilde F_3(z, \gamma) \equiv \frac{ (\chi+z q g) \xi^\gamma}{(1-z) \qty[1-z(1-2 q)]}  \, , \label{eq:functions-to-invert}
\end{align}
where $\gamma$ is an integer.
In terms of the above functions, one may recast Eq.~\eqref{eq:S_z} as
\begin{align}
     \widetilde Q(n,z|n_0) 
&= f^{-|n_c\!-\!n_0|} \qty[ f^{\alpha/2} \widetilde F_1(z,|n\!-\!n_0|) - \widetilde F_1(z,|n\!-\!n_0| \!+\! \alpha)  +  q g \widetilde F_2(z,|n\!-\!n_0|\!+\!\alpha) 
+ \widetilde F_3(z,|n\!-\!n_0|\!+\!\alpha)  ] \, .
\label{eq:app-Q-zinF}
\end{align}
From Eqs.~\eqref{eq:functions-to-invert} and~\eqref{eq:app-chi-xi-chi-1}, we define a set of new functions $\widetilde f_i (z) \equiv \xi^{-\gamma} \widetilde F_i (z,\gamma)$:
\begin{align}
    \widetilde f_1 (z) = \frac{ 1}{\sqrt{( 1\!-\!z \rho_+ \!) (1\!-\!z \rho_- \!)} } , \quad 
    \widetilde f_2(z) = \frac{ z }{(1-z) \qty[1-z(1-2 q)]} , \quad 
    \widetilde f_3(z) = \frac{ \sqrt{( 1\!-\!z \rho_+ \!) (1\!-\!z \rho_- \!)}}{(1-z) \qty[1-z(1-2 q)]}  \, , \label{eq:functions-to-invert-smallf}
\end{align}
such that once the $z$-inversions of all $\widetilde f_i (z)$ and $\xi^\gamma$ are known, one may convolute them in time to obtain the $z$-inversion of $\widetilde F_i (z,\gamma)$ and hence obtain the time-dependent propagator $Q(n,t|n_0)$ using Eq.~\eqref{eq:app-Q-zinF}.
In the following, we  restrict ourselves to providing the $z$-inversions of $\widetilde f_i (z)$ and $\xi^\gamma$.

To proceed, let us denote the $z$-inversion of a function $\widetilde f(z)$ by ${\mathfrak{Z}^{-1}}[\widetilde f(z) ]$. 
We first consider the function $\widetilde f_1(z)$, which with the use of the generalized binomial series for a negative half--integer exponent, namely,  $(1-x)^{-1/2} = \sum_{n=0}^{\infty} \binom{-1/2}{n}(-x)^n = \sum_{n=0}^{\infty} \frac{1}{4^n} \binom{2n}{n} x^n$ for  $|x|<1$, may be written as
\begin{align}
\widetilde f_1(z) = \sum_{k=0}^{\infty} \frac{1}{4^k} \binom{2k}{k} \rho_+^k  z^k \sum_{m=0}^{\infty} \frac{1}{4^m} \binom{2m}{m}  \rho_-^{m} z^m .
\label{eq:Fz_product}
\end{align}
Reindexing the double sum by defining $t=k+m$ such that for a fixed $t$, $k$ may range from $0$ to $t$, with $m = t - k$ and collecting the coefficient of $z^t$ gives
\begin{align}
\widetilde f_1(z) 
&= \sum_{t=0}^{\infty} \left[ \sum_{k=0}^{t} \frac{1}{4^t} \binom{2k}{k}  \binom{2t-2k}{t-k} \rho_+^k \, \rho_-^{t-k} \right] z^t ,
\label{eq:Fz_coeff}
\end{align}
which readily gives the $z$-inversion
\begin{equation}
{\mathfrak{Z}^{-1}}[\widetilde f_1(z) ] =f_1(t) 
=  \frac{1}{4^t} \sum_{k=0}^{t}  \binom{2k}{k}  \binom{2t-2k}{t-k} \rho_+^k \, \rho_-^{t-k} .
\label{eq:f1t_final}
\end{equation}

The $z$-inversion of $\widetilde{f}_2(z)$ is the simplest one. One may rewrite $\widetilde{f}_2(z)$
as a partial fraction
\begin{align}
\widetilde{f}_2(z) =  \frac{1}{2 q} \qty(\frac{1}{1-z}-\frac{1}{1-(1-2 q) z}) ,
\end{align}
and use the standard inverse transforms
$\mathfrak{Z}^{-1}\! [ \frac{1}{1-z} ] = 1$ and  $\mathfrak{Z}^{-1}\![ \frac{1}{1 - a z} ] = a^t $ to obtain
\begin{equation}
{\mathfrak{Z}^{-1}}[\widetilde f_2(z) ] =f_2(t)
=\frac{1 - \bigl(1 - 2q\bigr)^{t}}{2q}.
\label{eq:f2t-final}
\end{equation}

Next we turn to $\widetilde f_3(z)$ and first consider the factor
\begin{equation}
\frac{1}{(1-z)\,[1 - z(1-2q)]} = z^{-1} \widetilde f_2(z) = \sum_{t=0}^{\infty}f_2(t)z^{t-1} = \sum_{t=0}^{\infty}f_2(t+1)z^{t} .
\label{eq:factor-hl-series}
\end{equation}
We then expand the numerator $\sqrt{1-z \rho_+}\,\sqrt{1- z \rho_-}$ using the generalized binomial series to obtain
\begin{align}
\sqrt{1 - \rho_+ z} \sqrt{1 - \rho_- z}
&= \sum_{k=0}^{\infty} \binom{\tfrac12}{k}(-\rho_+)^k z^k 
\sum_{m=0}^{\infty} \binom{\tfrac12}{m}(-\rho_-)^m z^m \equiv
\sum_{k=0}^{\infty} A_k z^k,
\label{eq:numer-Ak-series}
\end{align}
with coefficients
\begin{equation}
A_k = \sum_{m=0}^{k} (-1)^k
\binom{\tfrac12}{m}\binom{\tfrac12}{k-m} \, \rho_+^{m} \rho_-^{k-m}.
\label{eq:Ak_def}
\end{equation}
Using Eqs.~\eqref{eq:factor-hl-series} and~\eqref{eq:numer-Ak-series}, we obtain 
\begin{equation}
\widetilde f_3(z) = \sum_{k=0}^{\infty} A_k z^k  \sum_{\ell=0}^{\infty} f_2(\ell+1) z^{\ell} 
= \sum_{t=0}^{\infty}
\left[
\sum_{k=0}^{t} A_k\,f_2(t-k+1)
\right] z^t ,
\end{equation}
which along with Eqs.~\eqref{eq:f2t-final} and~\eqref{eq:Ak_def} gives the inversion of $\widetilde f_3(z)$ as
\begin{align}
{\mathfrak{Z}^{-1}}[\widetilde f_3(z) ] =f_3(t)
= \frac{1}{2q}
 \sum_{k=0}^{t} \, \sum_{m=0}^{k} (-1)^k
\binom{\tfrac12}{m}\binom{\tfrac12}{k-m} \, \rho_+^{m} \rho_-^{k-m}  \,
\Bigl[1 - (1-2q)^{\,t-k+1}\Bigr] .
\label{eq:f3_final}
\end{align}

Finally, we tackle the inversion of $\xi^\gamma$ with $\gamma$ being integer. Using Binomial expansion, we write from Eq.~\eqref{eq:app-chi-xi-chi-1} that
\begin{align}
\xi^{\gamma}(z)&= \frac{1}{q^{\gamma}(1+g)^{\gamma} z^{\gamma}} \sum_{r=0}^{\gamma} \binom{\gamma}{r}  \big[1-z(1-q)\big]^{\gamma-r} (-1)^r \left[1-z \rho_+ \right]^{\frac{r}{2}} \left[1-z \rho_- \right]^{\frac{r}{2}} .
\label{eq:xi-to-gamma-app1}
\end{align}
Similar to Eqs.~\eqref{eq:numer-Ak-series} and~\eqref{eq:Ak_def}, we may write
\begin{align}
(1 - z \rho_+ )^{r/2} (1 - z \rho_- )^{r/2}
= \sum_{k=0}^{\infty} \sum_{m=0}^{k} (-1)^k
\binom{r/2}{m}\binom{r/2}{k-m} \, \rho_+^{m} \rho_-^{k-m} z^k,
\label{eq:numer-Ak-r2-series}
\end{align}
which when substituted in Eq.~\eqref{eq:xi-to-gamma-app1} yields
\begin{align}
\xi^{\gamma}(z) &= \frac{1}{q^{\gamma}(1+g)^{\gamma} z^{\gamma} }\sum_{r=0}^{\gamma}  \binom{\gamma}{r}  \sum_{s=0}^{\gamma-r} \binom{\gamma-r}{s}   z^{s} (1-q)^{s} (-1)^{r+s}  \sum_{k=0}^{\infty} \sum_{m=0}^{k} (-1)^k
\binom{r/2}{m}\binom{r/2}{k-m} \, \rho_+^{m} \rho_-^{k-m} z^k .
\label{eq:xi-to-gamma-app2}
\end{align}

Let us recall the formal definition of the $z$-inversion, namely, 
\begin{align}
{\mathfrak Z}^{-1} [{\widetilde{f}}(z)] \equiv \frac{1}{2\pi i} \oint \dd{z} ~{\widetilde{f}}(z) ~ z^{-t-1} ; \quad |z|<1, \label{eq:z-inversion}
\end{align}
where the integration contour is to be taken counter-clockwise. 
Considering the inversion of $\xi^{\gamma}(z)$ from Eq.~\eqref{eq:xi-to-gamma-app2}, we see that for $|z|<1$ the integrand $ \xi^{\gamma}(z) z^{-t-1}$ in~\eqref{eq:z-inversion} has only a single pole of order $(t+1+\gamma-s-k)$ at $z=0$. Choosing any closed contour with $0<|z|<1$, one may then evaluate the complex integral, which yields the residue $b_{-1}$ of $\xi^{\gamma}(z) z^{-t-1}$ at $z=0$. Using Eq.~\eqref{eq:xi-to-gamma-app2}, we expand $\xi^{\gamma}(z) z^{-t-1}$ as a Laurent series about the singularity $z=0$:
\begin{align}
\xi^{\gamma}(z) z^{-t-1} &= \frac{1}{q^{\gamma}(1+g)^{\gamma}  }\sum_{r=0}^{\gamma}  \binom{\gamma}{r}  \sum_{s=0}^{\gamma-r} \binom{\gamma-r}{s}    (1-q)^{s} (-1)^{r+s}  \sum_{k=0}^{\infty} \sum_{m=0}^{k} (-1)^k
\binom{r/2}{m}\binom{r/2}{k-m} \, \rho_+^{m} \rho_-^{k-m} z^{k+s-\gamma-t-1} .
\label{eq:xi-to-gamma-app3} 
\end{align} 
The coefficient of $z^{-1}$ on the right hand side of the above equation is the residue $b_{-1}$, which is obtained by setting $k=t+\gamma-s$ and removing the summation over $k$. We thus get 
\begin{align}
{\mathfrak Z}^{-1}[\xi^{\gamma}(z)] &= \frac{(-1)^{t+\gamma}}{q^{\gamma}(1+g)^{\gamma}  }\sum_{r=0}^{\gamma} (-1)^{r}  \binom{\gamma}{r}  \sum_{s=0}^{\gamma-r} \binom{\gamma-r}{s}    (1-q)^{s}   \sum_{m=0}^{t+\gamma-s} 
\binom{r/2}{m}\binom{r/2}{t+\gamma-s-m} \, \rho_+^{m} \rho_-^{t+\gamma-s-m}  .
\label{eq:xi-to-gamma-app4} 
\end{align}
Using Eqs.~\eqref{eq:f1t_final},~\eqref{eq:f2t-final},~\eqref{eq:f3_final}, and~\eqref{eq:xi-to-gamma-app4},
one obtains the $z$-inversion of all $\widetilde F_i (z,\gamma)$ in Eq.~\eqref{eq:functions-to-invert} as convolutions and then find the time-dependent propagator $Q(n,t|n_0)$ from Eq.~\eqref{eq:app-Q-zinF}.

\section{Steady state for bounded propagator in V--shaped confinement}
\label{sec:Vpot-refl-prop-app}

To find the propagator in the V--shaped confinement centred at $n_c=R$ with reflecting boundaries at 0 and $2R$, we proceed in two steps, first by finding the left bounded domain and subsequently the propagator for the fully bounded domain.  
Considering the domain bounded to the left a reflecting boundary at $n=0$, one may employ the same technique as in appendix~\ref{sec:Vpot-prop-app} and expand the left-bounded propagator $\widetilde L(n, z | n_0)$ from Eq.~\eqref{eq:Vpot-left-bound-prop} into series of $(1-z)$ to  obtain
\begin{align}
\widetilde L(n, z | n_0) \approx \frac{{L_0}}{1-z}+{L_1}+{L_2} (1-z)+{L_3} (1-z)^2 + \mathcal O(1-z)^3,
\label{eq:Ltilde-pole}
\end{align}
where one has
\begin{align}
    L_0 &\equiv \frac{2 g f^{| n-n_c| }}{2-(1-g) f^{n_c}} .
\end{align}
The remaining coefficients $L_1$, $L_2$, and $L_3$ can be computed, but do not provide additional analytical insight and have been omitted.
The steady state associated with the left-bounded propagator $\widetilde L(n,z|n_0)$ reads 
$L^{\rm ss}(n) \equiv  \lim_{z\to1}(1-z)\widetilde L(n,z| n_0)$ and yields Eq.~\eqref{eq:Vpot-semi-ss} of the main text.

Using the asymptotic form in (\ref{eq:Ltilde-pole}) with the $L_j$ coefficients for the semi-bounded propagator $\widetilde L(n, z | n_0)$,
we expand  $\widetilde Q^{\mathrm{ref}} (n, z | n_0)$ from Eq.~\eqref{eq:Vpot-fully-bound-prop} into a series of $(1-z)$ for $n_c=R$ as
\begin{align}
    \widetilde Q^{\mathrm{ref}}(n, z | n_0) \approx \frac{{A_0}}{1-z}+{A_1}+{A_2} (1-z)+{A_3} (1-z)^2 + \mathcal O(1-z)^3,
\label{eq:Qrtilde-pole}
\end{align}
where 
\begin{align}
   A_0 = \frac{g f^{| n-R| }}{  1- (1-g) f^R }.
   \label{eq:A0-def-app}
\end{align}
Given the cumbersome nature and excessive length of the form of the remaining coefficients $A_1$, $A_2$, and $A_3$, we have omitted them. 
The steady-state probability in this fully-bounded domain is obtained using   
$\widetilde Q^{\mathrm{ref}}_{\rm ss}(n) \equiv  \lim_{z\to1} (1-z)  \widetilde Q^{\mathrm{ref}}(n,z| n_0)$ and is reported in Eq.~\eqref{eq:Vpot-refl-full-ss} of the main text.

As the  expansion coefficients in Eq.~\eqref{eq:Qrtilde-pole} are in general functions of $n$ and $n_0$, i.e., $A_j = A_j(n,n_0)$, one may exploit the asymptotic expansion to compute the moments of the first-passage time from $n_0$ to $n$ in the fully-bounded domain. 
The generating function of the first-passage time  is simply given by
$\widetilde F^{\mathrm{ref}}(n, z | n_0) = \widetilde Q^{\mathrm{ref}}(n, z | n_0) / \widetilde Q^{\mathrm{ref}}(n, z | n)$, which when expanded into a series of $(1-z)$ using Eq.~\eqref{eq:Qrtilde-pole} produces
\begin{align}
    \widetilde F^{\mathrm{ref}}(n,z|n_0) \approx  B_0(n,n_0) + B_1(n,n_0) (1-z) + B_2 (n,n_0) (1-z)^2  + \mathcal O(1-z)^3 ,
\end{align}
where one has
\begin{align}
    B_0(n,n_0) &\equiv \frac{A_0(n,n_0)}{A_0(n,n)} , \\
    B_1(n,n_0) &\equiv \frac{A_0(n,n) A_1(n,n_0)  - A_0(n,n_0) A_1(n,n)}{A^2_0(n,n)} , \label{eq:B1-def-app} \\
    B_2(n,n_0) &\equiv \frac{ A_0(n,n_0) [ A_1(n,n)^2-A_0(n,n) A_2(n,n) ] + A_0(n,n) [ A_0(n,n) A_2(n,n_0)-A_1(n,n) A_1(n,n_0) ]}{A_0(n,n)^3} .
\end{align}
The mean-first passage time from $n_0$ to $n$ is then given by
$\expval{T} = \lim_{z\to 1} \partial \widetilde F^{\mathrm{ref}}(n,z|n_0) / \partial z = - B_1(n,n_0) $. 
Similarly, the second moment and variance of the first-passage time read $ \expval*{T^2} = \expval{T} + \lim_{z\to 1} \partial^2 \widetilde F^{\mathrm{ref}}(n,z|n_0) / \partial z^2 = \expval{T} + 2 B_2(n,n_0)$ and ${\mathrm{Var}}(T) = \expval{T} (1-\expval{T}) + 2 B_2 (n,n_0)$, respectively.
Noting from Eq.~\eqref{eq:A0-def-app} that $A_0(n,n_0) = A_0(n)$, we then write using Eq.~\eqref{eq:B1-def-app} that 
\begin{align}
    \expval{T} = \frac{A_1(n,n) - A_1(n,n_0)}{A_0(n)} ,
\end{align}
which after some tedious algebraic manipulation gives Eq.~\eqref{eq:vpot-fully-mfpt} of the main text.

\section{Bounded propagator in elastic-U shaped confinement}
\label{app:prop_upot}

We derive the propagator generating function for the U--potential by generalizing  Kac's procedure~\cite{kac_random_1947} introducing the diffusivity $q$. We start by rewriting the corresponding master equation~\eqref{eq:Master_Upot} in the matrix form ${{\mathbb{P}}_{t+1}} = {\mathbb{A}}   {\mathbb{P}}_t$, so that one has
\begin{align}
    P(n,t+1) &= \sum_{m=0}^{2R} {A}(n, m) P(m,t) \, ,
\end{align}
where $P(n,t)$ and ${A}(n, m)$ denote the element of the occupation probability vector ${\mathbb{P}}_{t}$ at time $t$ and the transition matrix $\mathbb{A}$, respectively. The quantity ${A}(n, m)$ sets the transition probability at each time step from site $m$ to site $n$. 
We enforce the reflecting boundary conditions at sites $0$ and $2R$.
Consequently, the relaxation matrix $\mathbb{A}_{(2R+1) \times (2R+1)}$ is given by
\begin{align}
    \mathbb{A} = 
     \begin{bmatrix}
        1-q & \frac{q}{2R}  & 0 & 0 & \cdots & 0 & 0 \\[1ex]
        q  & 1-q & \frac{2 q}{2R}  & 0 & \cdots & 0 & 0 \\[1ex]
        0 & q\qty(1-\frac{1}{2R}) & 1-q & \frac{3q}{2R}  &  \cdots & 0 & 0 \\[1ex]
        \vdots & \vdots & \ddots & \ddots & \ddots &  \vdots &  \vdots  \\[1ex]
        0 & 0 & \cdots  & \frac{3q}{2R} & 1-q & q\qty(1-\frac{1}{2R}) & 0 \\[1ex]
        0 & 0 & \cdots & 0  & \frac{2q}{2R}& 1-q & q \\[1ex]
        0 & 0 & \cdots & 0 & 0  & \frac{q}{2R}& 1-q  
     \end{bmatrix} \!\! .
\end{align}
Let us define $\mathbb{\Lambda_R}$ as the matrix, whose $j$th column is the $j$th right eigenvector of $\mathbb{A}$. 
Similarly, we define $\mathbb{\Lambda_L}$ as the matrix, whose $j$th row is the $j$th left eigenvector of $\mathbb{A}$. One may then write $\mathbb{A} = \mathbb{\Lambda_R} {\mathbb{D}} \mathbb{\Lambda_L}$ and $\mathbb{\Lambda_R} \mathbb{\Lambda_L} = {\mathbb{I}} = \mathbb{\Lambda_L} \mathbb{\Lambda_R} $,
where ${\mathbb{D}}$ is the diagonal matrix consisting of the $(2R+1)$ eigenvalues of $\mathbb{A}$ with ${\mathbb{D}}_{ij} = \lambda_j \delta_{i,j}$ with $\lambda_j$ being the $j$th eigenvalue, and $\mathbb{I}$ denotes the identity matrix.
The orthonormality condition satisfied by the matrices $\mathbb{\Lambda_R}$ and $\mathbb{\Lambda_L}$ is written as
\begin{align}
    \sum_{j=0}^{2R} \mathbb{\Lambda_R}^{ij} \mathbb{\Lambda_L}^{jk} = \delta_{i,k} \, . \label{eq:orthonormal}
\end{align}
The occupation probability vector ${{\mathbb{P}}_t}$ at time $t$ is obtained as
\begin{align}
    {{\mathbb{P}}_t} = {\mathbb{A}}\!^t \,  {\mathbb{P}}_0 = \mathbb{\Lambda_R} \, {\mathbb{D}}^t  \mathbb{\Lambda_L} {\mathbb{P}}_0 \, , \label{eq:Pt_sol}
\end{align}
where ${\mathbb{P}}_0$ is the initial probability vector.

In order to find the eigenvalue $\lambda$ and the eigenvector $(x_0, x_1, \ldots, x_{2R})^T$ of matrix $\mathbb{A}$, one needs to solve the following set of $(2R+1)$ simultaneous equations
\begin{align}
    \begin{matrix}
        0 &+ & (1-q) x_0 &+& \dfrac{q}{2R} x_1 & = \lambda x_0, \\[1.5ex]
        qx_0 &+& (1-q) x_1 &+& \dfrac{2q}{2R} x_2 & = \lambda x_1, \\[1.5ex]
        q\qty(1-\dfrac{1}{2R}) x_1 &+& (1-q) x_2 &+& \dfrac{3q}{2R} x_3 & = \lambda x_2, \\
        & & \vdots &  \\
        \dfrac{q}{2R} x_{2R-1} &+& (1-q) x_{2R} &+& 0 & = \lambda x_{2R} .
    \end{matrix} 
    \label{eq:simul_eqs}
\end{align}
To proceed, we first modify the last equation of the set~\eqref{eq:simul_eqs} and add an infinite number of auxiliary equations with unknown variables $x_{2R+1}, x_{2R+2}, \ldots$, to obtain an extended set of simultaneous equations:

\begin{align}
    \begin{matrix}
    0 & + & (1-q) x_0 &+& \dfrac{q}{2R} x_1 & = \lambda x_0 \\[2ex]
    qx_0 &+& (1-q) x_1 &+& \dfrac{2q}{2R} x_2 & = \lambda x_1 \\[2ex]
    q\qty(1-\dfrac{1}{2R}) x_1 &+& (1-q) x_2 &+& \dfrac{3q}{2R} x_3 & = \lambda x_2 \\
    & & \vdots & & & \\
    \dfrac{q}{2R} x_{2R-1} &+& (1-q) x_{2R} &+& q  \dfrac{2R + 1}{2R} x_{2R+1} &  \hskip6pt=\lambda x_{2R} \\[2ex]
    0 &+& (1-q) x_{2R+1} &+& q  \dfrac{2R + 2}{2R} x_{2R+2} & \hskip16pt= \lambda x_{2R+1} \\[2ex]
    -\dfrac{q}{2R} x_{2R+1} &+& (1-q) x_{2R+2} &+& q  \dfrac{2R + 3}{2R} x_{2R+3} & \hskip16pt= \lambda x_{2R+2} \\[2ex]
    & & \vdots &   .
    \end{matrix}
    \label{eq:simul_eqs_extended}
\end{align}
Note that the solutions for $x_0, x_1, \ldots, x_{2R}$ obtained from the  set~\eqref{eq:simul_eqs_extended}  reduce to the solutions of set~\eqref{eq:simul_eqs} when $x_{2R+1} = 0$.
Multiplying the equations of the set~\eqref{eq:simul_eqs_extended}, respectively, by $1,z,z^2,\ldots$, and summing them together, one obtains
\begin{align}
     \sum_{k=0}^{\infty} \qty(1-\dfrac{k}{2R}) x_k z^{k+1} + \sum_{k=0}^{\infty} \dfrac{ k}{2R} x_k z^{k-1}  = \dfrac{\lambda -1 + q}{q} \sum_{k=0}^{\infty} x_k z^k \, .
     \label{eq:sum_of_infinite_set}
\end{align}
Defining a generating function $\widetilde{f}(z) \equiv \sum_{k=0}^{\infty} x_k z^k$, Eq.~\eqref{eq:sum_of_infinite_set} may be recast in terms of $\widetilde{f}(z)$ as the following differential equation
\begin{align}
    \dv{\widetilde{f}(z)}{z} = 2R \frac{\lambda - 1 + q -z q}{q(1-z^2)} \widetilde{f}(z) \, ,
\end{align}
which with the condition $\widetilde{f}(0) = x_0 $ yields the solution
\begin{align}
    \widetilde{f}(z) =x_0 (1-z)^{\frac{R}{q} \qty(1-\lambda)} (1+z)^{\frac{R}{q} \qty(2q -1 + \lambda)} \, . \label{eq:ftilde_sol}
\end{align}
Note that choosing the values of $\lambda$'s as
\begin{align}
    \lambda_j = j \frac{q}{R} + 1 -2 q \, ; \quad j \in [0,1,2,\ldots,2R], \label{eq:eval}
\end{align}
we observe that the function $\widetilde{f}(z)$ is  a polynomial of degree $2R$ and hence we get $x_{2R+1} = 0$.
From Eq.~\eqref{eq:ftilde_sol}, we may then write
\begin{align}
    \widetilde{f}(z)   \equiv \sum_{k=0}^{2R} C^{(j)}_k z^k =  (1-z)^{2R-j} (1+z)^{j} \, ,
    \label{eq:C_jk_def}
\end{align}
with the choices
$x_0 = C^{(j)}_{0} = 1$, $x_{1\leq k \leq 2R} = C^{(j)}_{k}$, and $x_{k>2R} = 0 $.
As a result, the $k$th component of the $j$th right  eigenvector (column vector) of matrix $\mathbb{A}$ is identified as $C^{(j)}_k$, and therefore, we have
\begin{align}
    \mathbb{\Lambda_R} = \begin{bmatrix}
        C^{(0)}_0 & C^{(1)}_0 & C^{(2)}_0 & \ldots & C^{(2R)}_{0} \\[1ex]
        C^{(0)}_1 & C^{(1)}_1 & C^{(2)}_1 & \ldots & C^{(2R)}_{1} \\
        \vdots & \vdots & \vdots & \ddots & \vdots  \\[1ex]
        C^{(0)}_{2R} & C^{(1)}_{2R} & C^{(2)}_{2R} & \ldots & C^{(2R)}_{2R} 
    \end{bmatrix} .
\end{align}
Expanding Eq.~\eqref{eq:C_jk_def} in a power series of $z$ and computing  corresponding coefficients yield
\begin{align}
   \mathbb{\Lambda_R}^{kj} = C^{(j)}_k = \sum_{m=0}^{k} (-1)^m \binom{2R-j}{m} \binom{j}{k-m} \label{eq:Cjk_sum_app} , 
\end{align}
where one has $0\leq k,j\leq 2R$.
Equations~\eqref{eq:eval} and~\eqref{eq:Cjk_sum_app} give the eigenvalues and the components of the right eigenvectors of the matrix $\mathbb{A}$, respectively. 
Using the eigenvalues~\eqref{eq:eval}, one may write the components of the diagonal matrix ${\mathbb{D}}^t$ as
\begin{align}
    ({\mathbb{D}}^t)^{ij} = \qty(1-2 q + i \frac{q}{R})^t  \delta_{i,j} \ .
\end{align}

Following a similar approach, one may also obtain the elements of the matrix $\mathbb{\Lambda_L}$. Writing its elements as $\mathbb{\Lambda_L}^{ij} = E_{(i)}^j$, such that  $E_{(i)}^j$ is the $j$th component of the $i$th left eigenvector (row vector) of $\mathbb{A}$,
and using the orthonormality condition~\eqref{eq:orthonormal}, we identify 
\begin{align}
    \sum_{j=0}^{2R} C^{(j)}_i  E_{(j)}^k = \delta_{i,k} \, .
    \label{eq:E_jk_def}
\end{align}
Multiplying Eq.~\eqref{eq:E_jk_def} by $z^i$, then summing over $i$, and finally using Eq.~\eqref{eq:C_jk_def}, one obtains
\begin{align}
    \sum_{j=0}^{2R} E_{(j)}^k \qty(\frac{1+z}{1-z})^j = \frac{z^k}{(1-z)^{2R} } \, ,
\end{align}
which with the definition
\begin{align}
    \zeta \equiv \frac{1+z}{1-z}\, ,\quad z = - \frac{1-\zeta}{1+\zeta} \, , \quad 1-z = \frac{2}{1+\zeta} \, ,
\end{align}
yields
\begin{align}
    \sum_{j=0}^{2R} E_{(j)}^k \zeta^j = \frac{(-1)^k}{2^{2R} } \qty(1-\zeta)^k \qty(1+\zeta)^{2R-k} .
\end{align}
Expanding the right hand side of the above equation in powers of $\zeta$ and comparing the coefficients, we obtain
\begin{align}
    \mathbb{\Lambda_L}^{jk} = E_{(j)}^k &= \frac{(-1)^k}{2^{2R} } \sum_{m=0}^{j} (-1)^{j-m} \binom{2R-k}{m} \binom{k}{j-m} = \frac{(-1)^{k+j}}{2^{2R} } C^{(k)}_j  \, .
\end{align}

Finally, using Eq.~\eqref{eq:Pt_sol} with the initial condition ${\mathbb{P}}_0^{k} = \delta_{k, n_0} $, we get the propagator, i.e., the probability for the walker to be at site $n$ at time $t$ while starting from site $n_0$ to be 
\begin{align}
    P(n,t|n_0) &= {\mathbb{P}}_{t}^{n} =  \sum_{i,j,k=0}^{2R}\mathbb{\Lambda_R}^{ni} \, ({\mathbb{D}}^t)^{ij}  \mathbb{\Lambda_L}^{jk} {\mathbb{P}}_0^k 
    =  \frac{1}{2^{2R} } \sum_{j=0}^{2R}   (-1)^{j+n_0} C^{(j)}_{n}  C^{(n_0)}_{j} \, \qty(1-2 q + j \frac{q}{R})^t    \,  ,
    \label{eq:prop_OU_app}
\end{align}
which is the same as Eq.~\eqref{eq:prop_OU} of main text.
In the limit $z \to 1$, Eq.~\eqref{eq:C_jk_def} yields the identity $\sum_{n=0}^{2R} C^{(j)}_{n} = 2^j \delta_{j,2R}$, using which we obtain from Eq.~\eqref{eq:prop_OU_app} that
$\sum_{n=0}^{2R} P(n,t|n_0) = (-1)^{n_0} C^{(n_0)}_{2R} = \sum_{m=0}^{2R} (-1)^{m+n_0} \binom{2R-n_0}{m} \binom{n_0}{2R-m} $.
Note that the first and second binomials on the right-hand side of the last relation yield a non-zero contribution only when $2R - n_0 \geq m$ and $n_0 \geq 2R - m$, respectively.
As a result, in the summation over $m$, only the term with $m=2R-n_0$ yields non-zero contribution, and consequently, we get $\sum_{n=0}^{2R} P(n,t|n_0) = 1$, which proves the normalization of the propagator.

In terms of the (binary) Krawtchouk polynomials $K_k(x)$ with parameter $2R$, defined by
\begin{equation}
K_k(x) \equiv \sum_{m=0}^{k} (-1)^m \binom{x}{m}\binom{2R-x}{k-m},
\quad 0\le k,x\le 2R ,
\label{eq:K-def-app}
\end{equation}
one may identify from Eq.~\eqref{eq:Cjk_sum_app} that $C_k^{(j)} = K_k(2R-j);~ 0\le k,j\le 2R$. Then, reindexing the sum over $j$ in Eq.~\eqref{eq:prop_OU_app} with $x=2R-j$, we obtain
\begin{align}
P(n,t| n_0)
&= \frac{1}{2^{2R}}
\sum_{x=0}^{2R}
(-1)^{x+n_0} 
K_n(x) K_{2R-x}(2R-n_0) \left(1 - x\,\frac{q}{R}\right)^{t}.
\label{eq:P-K-2}
\end{align}

Let us  denote the coefficient $a_k$ of $z^k$ in a series $G(z) = \sum_{k=0}^{2R} a_k z^k$  using a notation such that $a_k = [z^k] G(z)$. 
From the generating function identity of the Krawtchouk polynomials, namely, 
\begin{equation}
\sum_{k=0}^{2R} K_k(x)\,z^k
= (1-z)^x(1+z)^{2R-x} ,
\label{eq:Knj-GF}
\end{equation}
we may identify
\begin{equation}
K_x(n_0)
= [z^{x}] F(z) , \quad F(z) \equiv (1-z)^{n_0}(1+z)^{2R-n_0} .
\label{eq:Kx-n0-coef}
\end{equation}
We then have
\begin{align}
K_{2R-x}(2R-n_0)
&= [z^{2R-x}]\bigl[(1-z)^{2R-n_0}(1+z)^{n_0}\bigr]  
= [z^{2R-x}] F(-z) 
= (-1)^{2R-x} \times [z^{2R-x}] F(z).
\label{eq:K2R-x-coef}
\end{align}
Note that one has
$z^{2R}F(1/z)
=
z^{2R}(1-1/z)^{n_0}(1+1/z)^{2R-n_0}
= (z-1)^{n_0}(z+1)^{2R-n_0} = (-1)^{n_0}F(z)$. Additionally, as $F(z)$ is a polynomial of degree $2R$, we may write $[z^{2R-x}]F(z)
=
[z^x]\bigl[z^{2R}F(1/z)\bigr]
=
[z^x]\bigl[(-1)^{n_0}F(z)\bigr]
=
(-1)^{n_0} [z^x] F(z)= K_x(n_0)$, where we have used Eq.~\eqref{eq:Kx-n0-coef}. From Eq.~\eqref{eq:K2R-x-coef}, we then obtain $K_{2R-x}(2R-n_0) = (-1)^{x+n_0} K_{x}(n_0)$, which when substituted in Eq.~\eqref{eq:P-K-2} yields Eq.~\eqref{eq:prop_OU} of the main text.

\section{First–passage statistics in elastic confinement}
\label{sec:fpt-Upot-app}

We start by breaking the sum over $j$ in the expression of the propagator generating function $\widetilde P(n,z|n_0)$ in Eq.~\eqref{eq:prop_elastic_genZ}.  Note that the $j=0$ term produces a pole at $z=1$ and yields the stationary distribution $P^{\mathrm{ss}}(n)$ [Eq.~\eqref{eq:ss-Upot}], which for notational brevity is denoted here as $\pi(n)$.
Isolating the stationary pole, we may rewrite
\begin{align}
\widetilde P(z,n| n_0)  &= \frac{\pi(n)}{1-z} + \sum_{j=1}^{2R}\frac{a_j}{1-z\lambda_j} =\frac{\pi(n)}{1-z}\,N(z),
\label{eq:Pz-split}
\end{align}
where 
\begin{align}
    a_j \equiv \frac{1}{2^{2R}}K_n(j)\,K_j(n_0) , \qquad
    N(z) \equiv 1 +\sum_{j=1}^{2R} \frac{c_j (1-z)}{1-z\lambda_j}, \qquad c_j \equiv \frac{a_j} {\pi(n)} .
\end{align}
Similarly, one may write
$\widetilde P(z,n| n)=\sum_{j=0}^{2R} \frac{b_j}{1-z\lambda_j} = \frac{\pi(n)}{1-z}\,D(z)$ with 
\begin{align}
b_j \equiv \frac{1}{2^{2R}}K_n(j)\,K_j(n), \qquad
D(z) \equiv  1 +\sum_{j=1}^{2R} d_j\,\frac{1-z}{1-z\lambda_j}, \qquad
d_j \equiv \frac{b_j}{\pi(n)}.
\end{align}
Note that $N(z)$ implicitly depends on both $n$ and $n_0$, while $D(z)$ has an implicit dependence on $n$ only. Here, we have $N(1)=D(1)=1$.
The first–passage time generating function is given by  
\begin{equation}
\widetilde F(n,z| n_0) = \frac{\widetilde P(z,n| n_0)}{\widetilde P(z,n| n)}
= \frac{N(z)}{D(z)}.
\label{eq:Fz-def-app}
\end{equation}
with $\widetilde F(n,1|n_0)=1$ as required for its normalization.

To find the moments of the first-passage time, one need the derivatives of $N(z)$ and $D(z)$ at $z=1$. To this end, let us first define, for $j\ge 1$, the function 
$f_j(z) \equiv \frac{1-z}{1-z\lambda_j}$, so that one has $N(z) = 1 + \sum_{j=1}^{2R} c_j f_j(z)$ and $D(z) = 1 + \sum_{j=1}^{2R} d_j f_j(z)$. One then obtains 
\begin{equation}
f_j'(1)=
-\frac{1}{1-\lambda_j}, \quad f_j''(1)
=
-\,\frac{2\lambda_j}{(1-\lambda_j)^2},
\label{eq:fderivs}
\end{equation}
where the prime denotes the derivative with respect to $z$.
Defining the sums
\begin{align}
S_1^{(c)} \equiv \sum_{j=1}^{2R}\frac{c_j}{1-\lambda_j}, \qquad 
S_2^{(c)} \equiv \sum_{j=1}^{2R}\frac{c_j\lambda_j}{(1-\lambda_j)^2}, \qquad  
S_1^{(d)} \equiv \sum_{j=1}^{2R}\frac{d_j}{1-\lambda_j}, \qquad 
S_2^{(d)} \equiv \sum_{j=1}^{2R}\frac{d_j\lambda_j}{(1-\lambda_j)^2} ,
\label{eq:sums-def-app}
\end{align}
and using \eqref{eq:fderivs}, we get
\begin{align}
N'(1)= -\,S_1^{(c)}, \qquad 
N''(1)= -2\,S_2^{(c)}, \qquad
D'(1)= -\,S_1^{(d)}, \qquad  
D''(1)= -2\,S_2^{(d)}. 
\label{eq:N-D-derivs_at_z1}
\end{align}
Using Eq.~\eqref{eq:Fz-def-app} and $N(1)=D(1)=1$, we obtain the mean first–passage time 
$\expval{T} = \widetilde F'(n,1|n_0)=N'(1)-D'(1)
= S_1^{(d)} -S_1^{(c)}$,
which along with Eq.~\eqref{eq:sums-def-app} yields
\begin{align}
\expval{T}
&= \sum_{j=1}^{2R}
\frac{d_j-c_j}{1-\lambda_j} = \frac{R}{q}
\sum_{j=1}^{2R}
\frac{d_j-c_j}{j}
\end{align}
Substituting the expressions for $d_j$ and $c_j$ in the above equation, we get Eq.~\eqref{eq:MFPT_Upot} of the main text.

We now move to the variance of the first-passage time. We denote the logarithm of the first-passage time generating function $\widetilde F(n,z|n_0)$ as
\begin{align}
    g(z) \equiv \ln \widetilde F(z)=\ln N(z)-\ln D(z) , \label{eq:gz-def-app}
\end{align}
where for brevity we have dropped the arguments $n,n_0$ from $\widetilde F$. With $\widetilde F(1)=1$ and $\widetilde F'(1)= g'(1) = \expval{T}$, the second moments is given by
$\expval{T^2} = \widetilde F''(1)+ \widetilde F'(1) =
g''(1)+g'(1)+\bigl(g'(1)\bigr)^2 = g''(1)+ \expval{T} + \expval{T}^2$. Thus, the variance may be obtained using
\begin{align}
\mathrm{Var}(T) = \expval{T} + g''(1).
\label{eq:var_T_ou_app-def}
\end{align}
Now, we have
$(\ln N)'={N'}/{N}$,
$(\ln N)''={N''}/{N}-\left({N'}/{N}\right)^2$
and similar expressions for $D$. Using $N(1)=D(1)=1$, we then obtain from Eq.~\eqref{eq:gz-def-app} that
$g''(1) = N''(1)-(N'(1))^2
- D''(1)+ (D'(1))^2$, which with Eq.~\eqref{eq:N-D-derivs_at_z1}
gives
\begin{equation}
g''(1)
=
2\bigl(S_2^{(d)}-S_2^{(c)}\bigr)
+
\bigl(S_1^{(d)}\bigr)^2
-
\bigl(S_1^{(c)}\bigr)^2.
\label{eq:gsecond}
\end{equation}
Using Eqs.~\eqref{eq:gsecond} and~\eqref{eq:var_T_ou_app-def}, we obtain the variance
\begin{equation}
\mathrm{Var}(T)
= \expval{T} + 2\bigl(S_2^{(d)}-S_2^{(c)}\bigr)
+ \bigl(S_1^{(d)}\bigr)^2 - \bigl(S_1^{(c)}\bigr)^2.
\label{eq:variance-final}
\end{equation}
From Eq.~\eqref{eq:sums-def-app}, one has $S_2^{(c)} = \sum_{j=1}^{2R} c_j/(1-\lambda_j)^2 - S_1^{(c)}$ and similarly for $S_2^{(d)}$. Using these relations and substituting expressions of the sums from Eq.~\eqref{eq:sums-def-app} in Eq.~\eqref{eq:variance-final}, we get 
\begin{align}
\mathrm{Var}(T) = \expval{T} + 2 \sum_{j=1}^{2R} \frac{ d_j-c_j}{(1-\lambda_j)^2}  
+  \sum_{j=1}^{2R} \frac{ d_j-c_j}{1-\lambda_j}    \qty[ \sum_{j=1}^{2R} \frac{ d_j+c_j}{1-\lambda_j} -2] 
= \expval{T} \qty[ \sum_{j=1}^{2R} \frac{ d_j+c_j}{1-\lambda_j} -1] + 2 \sum_{j=1}^{2R} \frac{ d_j-c_j}{(1-\lambda_j)^2} .
\label{eq:variance-fpt-app-00}
\end{align}
Substituting the expressions for $d_j$, $c_j$, and $\lambda_j$ in the above equation, we finally obtain Eq.~\eqref{eq:VarFPT_Upot} of the main text.

\end{widetext}

\bibliographystyle{unsrt}  
\bibliography{library-lrw}

\end{document}